\documentclass[
reprint,
superscriptaddress,
nofootinbib,
 amsmath,
 amssymb,
 aps,
prd,
floatfix,
showkeys
]{revtex4-2}
\usepackage{longtable}
\usepackage{aas_macros}
\usepackage{graphicx}
\usepackage{dcolumn}
\usepackage{siunitx}
\usepackage{bm}
\DeclareSIUnit \year {yr}
\usepackage[normalem]{ulem}

\usepackage[dvipsnames]{xcolor}

\renewcommand*{\i}{\mathrm{i}}

\begin{document}

\preprint{APS/123-QED}

\title{Simulating FRB Morphologies and Coherent Phase Correlation Signatures from Multi-Plane Astrophysical Lensing}

\author{Zarif Kader}
  \affiliation{Department of Physics, McGill University, 3600 rue University, Montr\'eal, QC H3A 2T8, Canada}
  \affiliation{Trottier Space Institute, McGill University, 3550 rue University, Montr\'eal, QC H3A 2A7, Canada}
\author{Matt Dobbs}
  \affiliation{Department of Physics, McGill University, 3600 rue University, Montr\'eal, QC H3A 2T8, Canada}
  \affiliation{Trottier Space Institute, McGill University, 3550 rue University, Montr\'eal, QC H3A 2A7, Canada}
\author{Calvin Leung}
\affiliation{MIT Kavli Institute for Astrophysics and Space Research, Massachusetts Institute of Technology, 77 Massachusetts Ave, Cambridge, MA 02139, USA}
\affiliation{Department of Physics, Massachusetts Institute of Technology, 77 Massachusetts Ave, Cambridge, MA 02139, USA}
\affiliation{Department of Astronomy, University of California Berkeley, Berkeley, CA 94720, USA}
\affiliation{NHFP Einstein Fellow}
\author{Kiyoshi W. Masui}
\affiliation{MIT Kavli Institute for Astrophysics and Space Research, Massachusetts Institute of Technology, 77 Massachusetts Ave, Cambridge, MA 02139, USA}
\affiliation{Department of Physics, Massachusetts Institute of Technology, 77 Massachusetts Ave, Cambridge, MA 02139, USA}
\author{Mawson W. Sammons}
\affiliation{Department of Physics, McGill University, 3600 rue University, Montr\'eal, QC H3A 2T8, Canada}
\affiliation{Trottier Space Institute, McGill University, 3550 rue University, Montr\'eal, QC H3A 2A7, Canada}

\newcommand{\allacks}{M.D. is supported by a CRC Chair, NSERC Discovery Grant, CIFAR, and by the FRQNT Centre de Recherche en Astrophysique du Qu\'ebec (CRAQ). C. L. is supported by NASA through the NASA Hubble Fellowship grant HST-HF2-51536.001-A awarded by the Space Telescope Science Institute, which is operated by the Association of Universities for Research in Astronomy, Inc., under NASA contract NAS5-26555. K.W.M. holds the Adam J. Burgasser Chair in Astrophysics. M.W.S. acknowledges support from the Trottier Space Institute Fellowship program.
}

\date{\today}

\begin{abstract}
Fast Radio Bursts (FRBs), like pulsars, display radio emission from compact regions such that they can be treated as point sources. As this radiation propagates through space, they encounter sources of lensing such as a gravitational field of massive objects or inhomogeneous changes in the electron density of cold plasma. We have developed a simulation tool to generate these lensing morphologies through coherent propagation transfer functions generated by phase coherent geometric optics on a spatial grid. In the limit an FRB can be treated as a point source, the ray paths from the FRB to the observer are phase coherent. Each image will have a time delay and magnification that will alter the emitted frequency-temporal morphology of the FRB to that which is observed. The interference of these images could also decohere the observed phase properties of the images, affecting any phase related searches such as searching for the auto-correlation of the observed FRB voltage with other images in time. We present analytic test cases to demonstrate that the simulation can model qualitative properties. We provide example multi-plane lensing systems to show the capabilities of the simulation in modeling the lensed morphology of an FRB and observed phase coherence.
\end{abstract}


\maketitle

\section{Introduction} \label{sec:intro}
Fast radio bursts (FRBs) (see \textcite{Petroff2022} for a review) hold the potential to be used as probes on extragalactic scales. The $\sim$ millisecond duration of FRBs and their brightness temperature imply a coherent emission mechanism from a compact region \cite{Petroff2022}. In this manner FRBs are like pulsars \cite{cordes2019}. Some FRBs have been localized to extragalactic sources \cite{Tendulkar2017} and within our galaxy an FRB has been associated with a magnetar \cite{Bochenek2020}, linking a population of FRBs to coherent emission from compact regions. Based on early results from surveys like the Canadian Hydrogen Intensity Mapping Experiment (CHIME) \cite{CHIME2021}, we expect many thousands of FRBs to be reported over the next few years. As a cosmological probe, these objects hold immense potential as they can probe extragalactic scales, have statistical sampling power, and have the potential to be coherently lensed. In the context of FRBs, accurate modeling of propagation effects is needed to disentangle it from the intrinsic emission process of the FRB. This paper provides, for the first time, a fast method to simulate the phase coherent changes to the electric field imposed by an astrophysical phenomenon acting as a lens -- an important contribution to modeling the various morphologies of FRBs.

Radio telescopes have the ability to measure the phase of the electric field using voltage timestreams from radio antennas. A detection of a coherently lensed FRB is one where a radio telescope is able to measure the coherent phase delay between the lensed images of the intrinsic electric field, analogous to a radio correlator with a fixed phase delay between baselines. In this paper, coherence is defined as follows: Two paths from a point source to a point observer are considered to be coherent if the phase delay is preserved between paths. If coherence is preserved, a search for phase correlations in the voltage data from a radio telescope can be done by time-lag correlating the voltage signal of different burst signals, as was done in \citet{Kader2022}. This method can allow one to detect sub-solar to solar mass gravitational lenses\cite{Kader2022,Leung2022}. A phase coherent search of lensing is of particular interest to constrain compact dark matter on cosmological scales, such as that from dark compact objects like primordial black holes \cite{Katz2020,Eichler2017,Leung2022}, and search for FRBs microlensed by planets \cite{Jow2020}. 

Separate from gravitational lensing, the plasma FRBs propagate through can also act like a dispersive plasma lens. FRBs are, thus far, observed to be point-like coherent sources like most pulsars. For this reason, the propagation phenomenon observed for pulsars will be the same for FRBs. For example, coherent time-lag phase correlations due to propagation effects, specifically from dispersive plasma, are a known phenomenon for pulsars. An inhomogeneous plasma can act like a diffraction grating and create a diffraction pattern in the frequency spectra of pulsars, through the superposition of electric fields with different time delays. This is commonly referred to as diffractive scintillation \cite{Cordes1986}. Another commonly observed phenomenon is scattering \cite{Rickett1977}, where the dispersive plasma will angularly broaden a point source and temporally broaden the intrinsic burst morphology through the phase changes it induces on the electric field. 

Modeling these propagation effects for both gravitational and plasma lensing can require wave optics considerations rather than only geometric optics. Specifically, the phase of the propagated electric field needs to be known to superimpose the electric fields and model the phase coherence of the observed electric field. Simulating propagation in the wave optics regime requires solving highly oscillatory integrals (see section \ref{sec:theory}) which are conditionally convergent. In \citet{Feldbrugge2019} it was shown these integrals can be solved using Picard-Lefshectz theory and applied to both gravitational and plasma lensing. However, this method requires an analytic model for the lens while in this paper we will focus on any generic model for the lens, including lenses generated from stochastic processes. In \citet{Grillo2018}, the oscillatory integrals were evaluated through the use of Fast Fourier Transforms (FFTs). This method applies to generic lensing models if the grid resolution is sufficient. However, the computational cost required for this method makes stochastic lens models hard to evaluate. We require our simulation framework to be able to model any lens and evaluate the propagation in a rather computationally efficient manner.

The Eikonal limit is the regime between geometric optics and wave optics \cite{Jow2022} where images can form and phase effects are considered. Pulsar scintillometry has shown simulations in this regime can replicate the observed scintillation patterns \cite{Walker2004,Cordes2006,Coles2010,Reardon2020,Sprenger2022}. We will, similarly, work in this regime for this work. FRBs, except for one case to date \cite{Wu2024}, do not have the duty cycle to measure the scintillation evolution like pulsars. Propagation measurements must therefore be done using single bursts. In this work, we focus on that aspect and will model the phase coherence evolution of a single FRB burst as it is propagated through a lens.

Simulating propagation effects is further complicated by the possibility of having multiple propagation phenomena occur, a scenario that is quite possible for FRBs. FRBs will have contributions of plasma scattering from both the FRB host galaxy and the Milky Way \cite{Masui2015,Sammons2023}. Our simulation framework must, therefore, be capable of coherently modeling multiple propagation effects in any order of occurrence. Plasma lens have been seen for pulsars\cite{Backer2000,Brisken2010,Kerr2018,Sprenger2022,Zhu2023} and generally involve multi-plane propagation. Plasma lensed FRBs\cite{Cordes2017} would have similar considerations. Additionally, coherent gravitational lensing requires understanding the phase evolution as the FRB propagates through plasma \cite{Katz2020,Eichler2017,Kader2022,Leung2022}. The ability to model the propagation is also vital for measurements of cosmological parameters \cite{Wucknitz2021,Tsai2023}.

The simulation framework provided in this paper can allow for a better understanding of how to disentangle the propagation effects from the intrinsic FRB emission mechanism. Our formalism is derived in section \ref{sec:theory}. The propagation will generate a phase evolution that alters the origin electric field. If one is able to measure this intrinsic phase relation then it is possible to remove it. For example, suppose the duration of the FRB emission is intrinsically short such that it measures the entire scattering transfer function from intervening plasma much like an impulse response. Then, for similar repeat bursts having the same transfer function, the measured transfer function of the first burst could be used to remove the scattering of the repeat bursts. For pulsars, it has been shown in \citet{Main2017} that giant pulses can be used to remove the interstellar scattering up to a decoherence time. The emission mechanism of the FRB is not known presently so measurements of propagation effects are needed to constrain the models. In this paper, we consider the FRB to be a point source. Scintillation will not occur when sources are extended. By measuring the scintillation in the host galaxy, one can put constraints on the emission mechanism based on the scintillation properties observed in an FRB\cite{Kumar2024}. Trying to extract scintillation from the host galaxy is difficult due to contributions from the Milky Way but with this simulation framework, the morphological properties of multiple scattering screens can be modeled and investigated.

The overall content of this paper is to provide an overview of the types of lensing from astrophysical sources relevant to FRBs, present the simulation framework to model coherent propagation of the electric field, and model select lensing systems that should be relevant to FRBs and highlight the associated morphological and phase signatures of these systems. In section \ref{sec:theory}, we begin with an overview of the relevant physics associated with radio wave propagation, the major assumptions taken in this paper, and the formalism used in the simulation. In section \ref{sec:observables}, we outline the observables relevant for radio telescopes, defining the formalism for baseband data and the time-lag correlation search for phase correlations. We also include a definition of what we consider as coherence with this formalism. In section \ref{sec:code}, we present the simulation logic, how the search algorithm detects images, how we simulate the radio baseband data, and how a time-lag phase correlation search is done in our test cases. In section \ref{sec:single}, we consider analytic test cases, providing examples of gravitational lensing from a point mass and plasma lensing from a refractive rational lens. In section \ref{sec:multi}, we present case studies of lensing systems relevant to FRBs such as a single scattering screen, dual scattering screens, a gravitational lens and a scattering screen, and a plasma lens and a scattering screen. 

\section{Formalism for Astrophysical Lensing} \label{sec:theory}
\begin{figure}[!htb]
    \centering
    \includegraphics[width=\columnwidth]{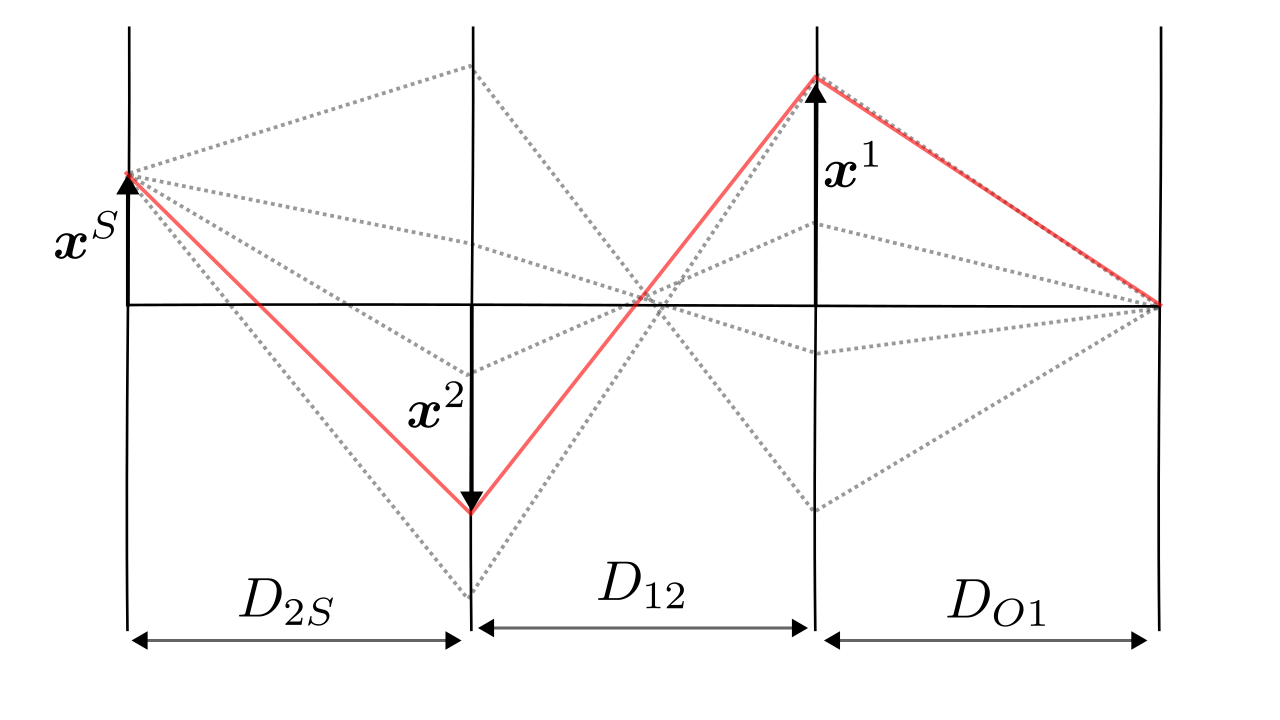}
    \caption{Illustrative diagram of a multi-plane lensing system. A point source on the source plane is found at position $\bm{x}^S$ where $\bm{x}^i$ is a vector spanning $\mathbb{R}^2$. There exist multiple thin screens with an angular diameter distance of $D_{ij}$ between the $i$-th and $j$-th plane. The Fermat potential is constructed as the delays for all possible paths (dashed gray lines) between the point source to the point observer set at the origin of the observer plane. The stationary paths (solid red line) of the Fermat potential are those that correspond to observed images.   }
    \label{fig:multiplane}
\end{figure}
The main motivation for the simulations developed in this paper is to model the propagation of radio waves through multiple screens, accounting for the phase evolution, to be able to evaluate the coherence between two paths. In this section, we detail the formalism that is used to create our simulation framework. We refer the reader to \citet{Nakamura1999}, \citet{Feldbrugge2019}, or \citet{Leung2023} for full derivations of lensing in wave optics regime and the Eikonal approximation. We will only summarize the Eikonal approximation and the assumptions needed for the approximation to be held.
\begin{enumerate}
    \item The wavelength is smaller than the distance traversed, $\lambda \ll D$, such that we have plane waves propagating along the optical axis directly to the observer from the source. The paraxial approximation can be applied to the wave equation in this scenario, $\nabla_\perp^2 E + 2 \i \frac{\omega}{c} \frac{\partial E}{\partial z} + \frac{\omega^2}{c^2} (n^2 - 1) E = 0 $ where the index of refraction is assumed to be spatially varying, $n(\bm{r},z)$. The solution to the electric field at any point in space is then evaluated by the Kirchoff-Fresnel diffraction integral (KDI) and the electric field will be of the form, $E(\bm{r},z,t) = E(\bm{r},z)e^{\i\omega t}$.
    \item The wavelengths are such that we are in the Eikonal limit, where $E(\bm{r},z) = e^{\i \omega T(\bm{r},z)}$ and the wave equation is given by the Eikonal equation, $ c |\nabla T| = n(\bm{r}) $. Here $T(\bm{r})$ is the Fermat potential. In this limit, we can apply the semi-classical approximation to the KDI. This is the regime between wave optics and geometric optics where rays are well defined and the wave effects of lensing will add a phase contribution to the electric field propagated along the rays. 
    \item The effective index of refraction in the system is a weak perturbation to a uniform index meaning the propagation can be represented as a thin phase screen. This is the limit in which the propagation contributions between two points occur at a distance that is small relative to the distance traversed by the wave. The index of refraction here is of the form $n(\bm{r}) = 1 + \phi(\bm{r})$  where  $0 \leq \phi(\bm{r}) \ll 1$ .
    \item The source of emission is point-like, i.e. pulsars and FRBs.
\end{enumerate}

Given the assumptions, let us consider a point on a source plane found at position $\bm{x}^{S}$. There exists a multitude of paths from this point source to a point observer at the origin of the observer plane. These paths are illustratively shown in figure \ref{fig:multiplane} as dashed lines originating from a point source.  Between the source and the observer there exist $N$ lensing planes which are numbered in reference to the observer plane and indexed by $i$ where $i=1,2,...,N$. The observer plane is indexed as $i=0=O$  and the source plane by $i=N+1=S$. Each plane is parameterized by a dimensionless vector,  $\bm{x}^{i}= x^{i}_a$, where $a=1,2$ and located at an angular diameter distance, $D_{Oi}$, from the observer. Given a refractive index, $n$, the Fermat potential can be constructed as all possible paths connecting the point source to the point observer,
\begin{equation}
    T=\int_{\mathrm{source}}^{\mathrm{observer}} n \frac{dl}{c}.
\end{equation}
A thin screen will add an instantaneous delay at the location of the screen while the regions between planes have $n=1$ corresponding to free-space propagation.

The multi-plane Fermat potential is given by \cite{Schneider1992, Feldbrugge2023, Blandford1986},
\begin{equation}\label{eq:fermfull}
    T(\bm{x}^1,\bm{x}^2,...,\bm{x}^{k},\bm{x}^{S},f) = \sum_{i=1}^{N}T^i(\bm{x}^i,\bm{x}^{i+1},f) ,
\end{equation}
where the $i$-th plane contribution is given by,
\begin{equation}\label{eq:fermpotential}
    \begin{split}
        T_i(\bm{x}^i,\bm{x}^{i+1},f) &= \mu_{g,i} \frac{(x^i_a - \frac{\eta_{i+1}}{\eta_{i}} x^{i+1}_a)^2}{2}\\
        &+ \mu_{l,i}(f) \Phi_i( \bm{x}_i) ~,\\
    \end{split}
\end{equation}
$\eta_i$ is some characteristic angular lensing scale defined for the $i$-th lens plane ($\frac{r_i}{D_{Oi}} = \theta_i = \eta_i x_i $). $\frac{\eta_{i+1}}{\eta_{i}}$ is the projection scaling of the previous plane into the next plane, where the projection of the source plane to the $N$-th plane is defined as $\frac{\eta_{S}}{\eta_{N}}=1$. The geometric parameter is defined as $ \mu_{g,i}= \frac{(1+z_i)D_{Oi}D_{Oi+1} \eta_{i}^2}{c D_{ii+1}}$ and $z_i$ is the redshift of the lens plane. We impose that our lensing potentials must have the frequency dependence of the lens be separable from the spatial model of the lens. With this assumption, we define a chromatic lensing parameter $\mu_{l,i}(f)$ and an achromatic lens potential, $\Phi_i(\bm{x}_i)$. This assumption holds if the ``type of lens" is constant across the entire thin screen such that the lens parameter has no spatial dependence, e.g. the plane is exclusively a gravitational lens or exclusively a cold, dispersive plasma lens. One final parameter definition is the frequency-dependent strength of the lens \cite{Clegg1998,Jow2022},
$ \alpha_i = \frac{\mu_{l,i}(f)}{\mu_{g,i}} $, which will be used for ease of computation later. 

For a frequency $f$, the electric field for a point observer at the origin of the optical axis of a multi-plane system is given by the Kirchoff-Fresnel diffraction integral \cite{Nakamura1999,Feldbrugge2023},
\begin{equation}\label{eq:fullmultiKDI}
\begin{split}
    E_S(\bm{x}^{S}, f) &\propto \int d\bm{x}^{1} d\bm{x}^{2}...d\bm{x}^{N}  E'(\bm{x}^{S}, f) \\
   & \times \exp\left(\i 2 \pi f T(\bm{x}^1,\bm{x}^2,...,\bm{x}^{N},\bm{x}^{S},f) \right) . \\
\end{split}
\end{equation}
$E'(\bm{x}^{S}, f)$ is the field value at a point at $\bm{x}^S$. A point source would be $E'(\bm{x}^{S}, f) = E'(f) \delta(\bm{x}^{S}- \bm{x'}^{S}) $. The time evolution of the field is given by, 
\begin{equation}\label{eq:timedom}
E_S(\bm{x}^{S}, t) = \int df E_S(\bm{x}^{S}, f) \exp\left(-\i 2 \pi f t \right),
\end{equation}
where the lensing potential and sources are assumed to be stationary. The procedure to evaluate the propagation with our simulation is to first evaluate $E_S(\bm{x}^{S}, f)$ and then perform a Fourier transform to obtain the electric field as a function of position on the source plane and time. Through equations \eqref{eq:fermfull} and \eqref{eq:fullmultiKDI}, it can be seen that the solution to the multi-plane KDI can be evaluated iteratively \cite{Feldbrugge2023}. For lensing, the normalization of the KDI is chosen such that we see the change to the electric field relative to the unlensed propagation, otherwise referred to as the amplification factor \cite{Nakamura1999,Leung2023}. The unlensed normalization is given by evaluating a Guassian integral \cite{Nakamura1999} which evaluates to $e^{-\i\frac{\pi}{2}} f \mu_{g,i}$.

Currently, all considerations have been in the wave optics regime. Unfortunately, the oscillatory nature of this integral makes it hard to efficiently evaluate numerically. They can require analytic continuation \cite{Feldbrugge2019} or computationally expensive Fourier transforms \cite{Grillo2018}. We can approximate the evaluation of this integral in the asymptotic limit $\frac{f T}{2 \pi} \gg 1$. This is the Eikonal limit\cite{Jow2022}. In this regime, we can approximate the contributions to the KDI as only those from points of stationary phase. These points are given by 
\begin{equation}\label{eq:gradpnt}
\begin{split}
    \frac{\partial T_i}{\partial x^{i}_a} = 0 = x^{i}_a - \frac{\eta_{i+1}}{\eta_{i}} x^{i+1}_a + \alpha_i \frac{\partial \Phi_i}{\partial x^{i}_a}.
\end{split}
\end{equation}

If the entire propagation occurs in the Eikonal limit, we can consider the full set of stationary paths to be the observed images of the point source to the point observer. Starting from the source plane with a point source at $\bm{x'}^S$, consider the set of solutions that satisfy the stationary condition on the $N$-th plane, $\{\bm{x}'^{N} | x'^{S}_a = x^{N}_a+ \alpha_N \partial^N_a \Phi_N \}$. These points themselves will behave like a point source such that, for each point in this set, there will exist another set of solutions for the stationary points on the next plane, $\{\bm{x}'^{N-1} | x'^{N}_a = \frac{\eta_{N-1}}{\eta_{N}} ( x^{N-1}_a+ \alpha_{N-1} \partial^{N-1}_a \Phi_{N-1}) \}$. Iterating until the $i=1$ plane we will have $N_l$ total paths where we define each path as $\gamma^l( \bm{x}'^{1},\bm{x}'^{2},...,\bm{x}'^{N}, \bm{x}'^{S})$ where $l=1,2,...,N_l$. In figure \ref{fig:multiplane}, one such path of all possible paths is highlighted as the red solid path. By finding all stationary (red) paths, the phase coherent images are found. For each plane, we perform a Taylor expansion about these points such that,
\begin{equation}\label{eq:taylorexp}
\begin{split}
    &T_i(\bm{x}^{i},\bm{x}^{i+1},f) \approx T_i(\gamma^l,f)\\
    &+ \frac{1}{2} \frac{\partial^2 T_i(\gamma^l,f)}{\partial x^{i}_{a} x^{i}_{b}} (x^{i}_{a} -  x^{i}_{a}(\gamma^l)) (x^{i}_{b} -  x^{i}_{b}(\gamma^l)),  \\
\end{split}
\end{equation}
where $\frac{\partial^2 T_i(\gamma^l,f)}{\partial x^{i}_{a} x^{i}_{b}}$ is the Hessian of the Fermat potential. In the limit where we start from a point source and propagate to a point observer and all lensing planes are in the Eikonal limit, there will exist a finite set of total image points generated by propagation. Further, suppose that all these stationary points have a non-zero determinant for the Hessian of the Fermat potential of each plane that makes all these points Morse \cite{Petters1992}. Given this, we can use Morse theory to describe the Fermat potential and construct a region around each stationary point that will be either a local minimum, maximum, or saddle point \cite{Petters1992}. If the region around one stationary point is well isolated from the region around another stationary point, then we can approximately evaluate the KDI integral as the contributions from the regions around these points. In the limit where we also have rapid phase oscillations, we can extend these regions to span the whole plane such that the phase oscillations will asymptotically go to zero in overlapping regions. The KDI integral around a single point is 
\begin{equation}\label{eq:iter_}
\begin{split}
    &E_{i+1}(\gamma^{l}, f)  = e^{-\i\frac{\pi}{2}} f \mu_{g,i} E_{i}(\gamma^l, f)  e^{\i2 \pi f \tau^l_i} \int d\bm{x}^{i}  \\
    & \exp\left(\i 2 \pi f \frac{1}{2} \frac{\partial^2 T_i(\gamma^l,f)}{\partial x^{i}_{a} x^{i}_{b}} (x^{i}_{a} -  x^{i}_{a}(\gamma^l)) (x^{i}_{b} -  x^{i}_{b}(\gamma^l))\right) ,\\
\end{split}
\end{equation}
where we define the delay imposed by the $i$-th lens for the $l$-th image as, 
\begin{equation}\label{eq:multidelay}
   \tau_i^{l} = T_i(\gamma^l,f) ~.
\end{equation}
The eigenvalues of the Hessian are,
\begin{equation}\label{eq:eigen}
\begin{split}
     \lambda^i_{1,2}(\gamma^{l}, f) &= \mu_{g,i}\bigg[ 1 + \frac{\alpha_i}{2}\bigg( \partial^i_{11}\Phi_i +\partial^i_{22}\Phi_i\\
    &\pm \sqrt{(\partial^i_{11}\Phi_i- \partial^i_{22}\Phi_i)^2 + 4 (\partial^i_{12}\Phi_i)^2} \bigg) \bigg],\\
\end{split}
\end{equation}
and evaluating the Gaussian integral\cite{Nakamura1999}, we obtain
\begin{equation}\label{eq:iter_func}
\begin{split}
    E_{i+1}(\gamma^{l}, f) & = \varepsilon^{l}_i e^{\i2 \pi f \tau^l_i} E_{i}(\gamma^{l}, f) ,
\end{split}
\end{equation}
where the magnification for a stationary point is defined as\cite{Nakamura1999,Leung2023},
\begin{equation}\label{eq:magni}
\begin{split}
    \varepsilon^{l}_i = \frac{\mu_{g,i}}{\sqrt{|\lambda^i_{1}| |\lambda^i_{2}}|} e^{\i\frac{\pi}{4} (sgn(\lambda^i_{1}) + sgn(\lambda^i_{2}) - 2)} .
\end{split}
\end{equation}

From equations \ref{eq:eigen} and \ref{eq:magni}, the magnification of an image is seen to depend on the lens strength $\alpha_i$, and the lens potential $\Phi_i$. The complex phase factor in equation \ref{eq:magni} is dependent solely on the curvature around the stationary point, where it will be $e^{\i0},e^{-\i\pi},e^{-\i\frac{\pi}{2}}$ for a local minimum, maximum, or saddle point, respectively. By determining the magnification from the eigenvalues of the Hessian of the Fermat potential, we can determine if the stationary point is at a caustic, which is identified by one of the eigenvalues being 0. This semi-classical approximation to the KDI won't be valid at these points or with points close to these caustics, i.e. Morse theory will not hold as the determinant of the Hessian is zero and stationary points are no longer isolated relative to another. In this case, our simulation will not be valid at or near caustic points. In these regimes, other methods are needed to evaluate the integral such as catastrophe theory \cite{Blandford1986} or Picard-Lefshectz theory \cite{Feldbrugge2019}. We will work throughout the paper in the Eikonal limit. Pulsar scintillometry has shown geometric optics is sufficient to model pulsar scintillation \cite{Walker2004,Cordes2006,Coles2010,Brisken2010,Reardon2020,Sprenger2022}. While the simulation toolset presented in this paper is not valid at caustics, we are interested in the ability to model a wide variety of lensing scenarios. Caustic lensing events are rarer in occurrence than non-caustic events as they require specific alignment of source and lens. This simulation framework will therefore be useful in modeling the majority of non-caustic lensing events and other simulations will be needed to model the select few caustic lensing events.

In the Eikonal limit, the final field for a path $l$ and image is given as,
\begin{equation}\label{eq:iter_final}
\begin{split}
    E(\gamma^{l}, f) & = E(\bm{x}'^S, f) \delta(\bm{x}^S - \bm{x}^S(\gamma^{l})) \prod_{i}^N \varepsilon^{l}_i e^{\i2 \pi f \tau^l_i} .
\end{split}
\end{equation}
We can define a propagation transfer function for the image path as 
\begin{equation}\label{eq:transferfunc}
\begin{split}
    H(\gamma^{l}, f) & = \bar{\varepsilon}^l e^{\i2 \pi f \bar{\tau}^l} \delta(\bm{x}^S - \bm{x}^S(\gamma^{l})),
\end{split}
\end{equation}
where the total magnification for the $l$-th image is defined as $\bar{\varepsilon}^l=\prod_i^{N} \varepsilon_i^l$, the total delay is defined as $\bar{\tau}^l=\sum_i^{N} \tau^l_i$ and the final image position on the source plane is given by $\delta(\bm{x}^S - \bm{x}^S(\gamma^{l}))$. These are the observables of lensing: the magnification $\bar{\varepsilon}^l$, the delay $\bar{\tau}^l$, and the image position $\bm{x}^S(\gamma^{l})$.

\subsection{Lensing Observables in Data}\label{sec:observables}
In the previous section, we showed how to obtain the electric field as observed by a point observer. In this section, we present how these observables will manifest in the data product of a radio telescope and how to search for the effects of coherent propagation through a time-lag correlation. 

We have a transfer function, equation \ref{eq:transferfunc}, such that using equations \ref{eq:timedom} and \ref{eq:iter_final} we can obtain the time dependent electric field as,
\begin{equation}\label{eq:timedomgeom}
E(\gamma^l, t) = \int df E(\bm{x}'^S, f) H(\gamma^{l}, f) e^{\i 2 \pi f t}.
\end{equation}
This propagated electric field is captured by a radio antenna and we consider the data in the voltage timestream recorded after being sent through a signal processing chain which includes a beam response, $B(\bm{\theta})$. The voltage timestream is then effectively represented by, 
\begin{equation}
    V(\bm{\theta},t) = \int d\bm{\theta}' B(\bm{\theta}'-\bm{\theta}) \sum_l^{N_l} E(\bm{\theta}'(\gamma^l), t) ~.
\end{equation}
If the beam response is common to all images, i.e. the instrument's beam width is significantly larger than the image separations, we have 
\begin{equation}
    V(t) = B \sum_{l}^{N_l} E(\gamma^l, t) ~.
\end{equation}
Conversely, if the beam can resolve image separations, 
\begin{equation*}
    V(\bm{\theta},t) = \sum_l^{N_l} B(\bm{\theta}(\gamma^l)) E(\bm{\theta}(\gamma^l), t) ~.
\end{equation*}
This is the first potential observable for lensing, the spatial separation between images in the sky. For this study, we are interested in lensing systems where the largest separation between images is significantly smaller than the width of the radio telescope beam. In this case, only the magnification and time delay of images are found to affect the field in the time and time-lag domain.

For radio telescopes, the main data product we will consider is baseband data. This is complex-valued frequency-temporal data representing the voltage timestream subdivided into various frequency channels, $V(f_k,t)$. If we start from a signal purely in the frequency domain, the time domain representation of a signal observed in a single frequency channel, centered at frequency $f_k$, is given by
\begin{equation}\label{eq:baseband}
    V(f_k,t) = \int_{- \infty}^{ \infty} df S(f) H(f) P(f- f_k ) e^{-\i 2 \pi f t} + N(f_k,t) ,
\end{equation} 
where $P(f-f_k)$ is the channel response, $S(f)= B E(\bm{x'}^S,f)$ is the voltage signal of the unlensed FRB emission, and $N(f_k,t)$ is the noise contribution of that channel. For the idealized telescope, one wants the response to be maximal near the channel center while the contributions from outside the channel width, i.e. spectral leakage, are reduced to a negligible level. The main importance of the channel response and equation \ref{eq:baseband} is that baseband data at a frequency $f_k$ is the integrated response over the channel width. This has implications for searching in the time-lag domain as this will alter expected phase correlations and the observed phase coherence of the transfer function. See section \ref{sec:coherence} for more details.

In the time domain, one will observe morphological alterations to the original FRB. In this paper, we model the original FRB as a simple Gaussian in the time-domain, $\left <|E(\bm{x}'^S,t)|^2\right > = A \exp\left(-\frac{(t - t_S)^2}{\sigma_S^2} \right)$, where $A$ is the intrinsic intensity, $t_S$ is the unlensed arrival time, and $\sigma_S$ is the temporal width of the burst. The left panels of figures \ref{fig:pm_dual} and \ref{fig:plasma_dual} show examples of these time-resolved images where the images of the original electric field will arrive at the observer at different times. Magnifications may change as a function of frequency as shown in the left panel of figure \ref{fig:plasma_dual} and if the arrival delays for images are small, they may be superimposed in time creating complex morphological structures like scattering tails which can be seen in the left panel of figure \ref{fig:gscatt_dual}.

While the time domain allows one to see the morphology, it is in the time-lag domain that one can detect lensing signatures through the search for phase coherent correlations of the electric field. For baseband data, this is done through a time-lag auto-correlation, i.e. for each observed frequency channel we can time-lag correlate the complex data per frequency channel. The expectation for time-lag auto-correlation of baseband is,
\begin{equation}\label{eq:corr}
\left <C(f_k,\hat{t})\right > = \int dt  \left <V(f_k,t)  V^*(f_k,t-\hat{t})\right > .
\end{equation}

For instrument data with noise present, one ideally wants to increase the signal weighting by applying a matched filter \cite{Kader2022}, which will result in the time-lag correlation function of a frequency channel given as,
\begin{equation}\label{eq:corre}
C'(f_k,\hat{t}) = \frac{ \int dt  V(f_k,t)  V^*(f_k,t-\hat{t}) |W(f_k,t)|^2 }{\sqrt{\int dt  |V(f_k,t)|^2  |V(f_k,t-\hat{t})|^2 |W(f_k,t)|^4} } .
\end{equation}
where the denominator is the normalization to the extra variance added from searching for phase correlations with a filter, $W(f_k,t)$. In this paper, we'll use this method to present the correlations where the filter is constructed by taking the power, $\sum_{f_k}|V(f_k,t)|^2$, smoothing with a Gaussian kernel, and clipping around the brightest burst (see \citet{Kader2022} for a more detailed description). Note that equation \ref{eq:corre} assumes the unlensed FRB emission is uncorrelated in phase at times larger than the reciprocal of the bandwidth, e.g. for CHIME we assume $\left <E(\bm{x}'^S,t)E^*(\bm{x}'^S,t-\hat{t})\right > = 0 ~, \tau > 1.25 ~\mathrm{ns}$.

As propagation transfer functions and, therefore, the time-lag correlation function are only analytic in very select cases, the correlation response is best illustrated numerically. One observable of key importance is whether the phase correlation is a chromatic or achromatic phase response. Plasma lensing is chromatic, whereas gravitational lensing is intrinsically achromatic. Therefore, any achromatic phase responses should be flagged as potential gravitational lensing images. Conversely, chromatic phase responses should be flagged as potential plasma lensed images. In our analytic test case, we present the correlation response of an achromatic time-lag correlation which can be seen in the right panel of figure \ref{fig:pm_dual}. For a frequency-dependent phase response we consider a test case of a Gaussian plasma lens, where the correlation signature can be seen in figure \ref{fig:plasma_dual}. 

In summary, there are three main observables for lensing; the spatial separation of images, the magnification of images, and the time delay of images. Assessing whether a specific morphology likely arises from propagation effects can be done through a time-lag correlation search for phase correlations. If the transfer function is coherent then the phase relation can allow one to determine if the morphology arises from propagation. Additionally, the chromatic nature of the phase response should allow one to determine whether one has observed gravitational or plasma lensing. If one has found a lensing signature, the next natural step might be to model the full lensing system. If the spatial separation isn't observable then even with the other two observables the key parameters of the lensing system can be determined. For example, the product of the mass of a point mass gravitational lens and the redshift of the lens can be uniquely determined from the delay between the two images and their relative magnification\cite{Kader2022}. For complicated morphologies and propagation through multiple screens, the new simulation toolset described here provides a method to model these systems and fit them to determine the lensing parameters.

\subsection{Phase Coherence of the Transfer Function}\label{sec:coherence}

One of the most important considerations for astrophysical lensing is whether coherence is maintained from the FRB to the observer. This is imperative for detecting a lensing signal using a time-lag correlation search. In this section, we discuss and consider what it means to have a coherent transfer function.  

Coherence in this paper is a statement about whether the FRB can be treated as a point source. If this assumption holds, then along the stationary paths connecting the point source to an observer, the Fermat potential is constant (per frequency) and the phase (or phase relation as a function of frequency) of the source is preserved. The smaller the transverse emission size of FRBs is, the more point-like and coherent the FRB is. To date, while methods have been suggested to determine this scale\cite{Kumar2024}, the transverse emission size is dependent on the model of the FRB emission and can vary in range. Note that diffractive scintillation, which is a phase coherent lensing effect, is seen in FRBs \cite{Schoen2021,Masui2015,Wu2024} suggesting FRBs can be point-like to a sufficient degree. Furthermore, in \citet{Kader2022}, the search for coherent phase correlations in the time-lag domain shows evidence for coherence being preserved along some sight lines. 

Suppose the FRB emission region extends to a scale $\bm{x}^{\zeta,S}$ from a point source at $\bm{x}'^{S}$. For coherent lensing to occur, the total transverse extent of the source up to this scale must not alter the Fermat potential to a measurable degree, i.e. the path delay from propagation between $\bm{x}^{\zeta,S}- \bm{x}'^{S}$ and  $\bm{x}'^{S}$ must be approximately the same. In this definition, the FRB emission region is ``unresolved'' by propagation and can be considered a point source rather than an extended source. To define this condition, we must consider when the stationary points are Morse and the region around each stationary point where the phase is not altered to a significant degree.

In the unlensed case, whether a source can be considered ``resolved" is set by the Fresnel scale, $\theta_F = \sqrt{\frac{c D_{ii+1}}{ f D_{Oi}D_{Oi+1}}}$, such that sources of approximately that scale will have the phase differences from the Fermat potential on the order of $\sim \pi$. For the unlensed case, this requires the FRB source region to be smaller than the Fresnel scale, $\theta_{\mathrm{FRB}}< \theta_F$ such that a completely incoherent source at that scale will have the same path delays from propagation. This condition must be satisfied for any coherent propagation to occur. 

Lensing adds more conditions that must be met. In the Eikonal limit, lensing creates image copies of the source at stationary points on the lens plane. For coherent propagation to occur, we require that the total propagation delays and magnifications are approximately constant in the regions surrounding stationary points. This requires that the total angular extent of the source, $\bm{x}^{\zeta,S}- \bm{x}'^{S}$, exist within a locally flat neighborhood around every stationary point. All points within this region will effectively have the same delay and magnification, i.e. equations \ref{eq:magni} and \ref{eq:multidelay} are applied to all points within this local neighborhood. By expressing the geometric delay parameter in terms of the Fresnel scale, $ \mu_{g,i}= \frac{(1+z_i) \eta_{i}^2}{f \theta_F^2}$, equation \ref{eq:eigen} transforms the Fresnel scale into the eigenspace around a stationary path originating from $\bm{x}'^S$. Under this formalism, the total angular extent of the source, $\bm{x}^{\zeta,S}- \bm{x}'^{S}$, projected in the eigenspace must not exceed the scale of the minimum eigenvalue, $f\lambda^i_{\mathrm{min}}$ else the phase will be altered by $> \pi$. In the regime where lensing itself is a stochastic process, such as scintillation, this scale is set by the diffractive scale \cite{Narayan1992}.

In the multi-plane regime, the general requirement is that the largest angular extent from the previous plane ( i.e. the largest separation between images) must be smaller than the smallest coherence scale where the phase can vary by $\sim \pi$. If this condition is true for all stationary paths, then the entire propagation transfer function is coherent. As stationary points can be relatively isolated, there does exist a semi-coherent regime where the largest angular extent of the previous plane is a point source around one image point but an extended source around another. In this regime, the phase coherence could be lost for some stationary paths but not all. In this case, the incoherent regime for lensing propagation is when the source is extended for all image paths.

Separate from an extended source, proper motion can also lead to a decoherence of the propagation. Proper motion becomes a problem if the source shifts over the minimum coherence scale resulting in a phase shift that alters the propagation coherence. We require that the entire lensing system must remain static. For coherence to be preserved, the proper motion must not shift the angular extent of the source over the coherence scale for the duration of the FRB burst, $\dot{\theta} < \frac{\theta_{\mathrm{coh}}}{\Delta t}$. If a lens has a significant proper motion that shifts the effective source position between this time scale, then the formalism presented in this paper cannot reconstruct this dynamical lens. In the multi-plane lensing scenario, it would be possible to have two images generated by a previous lens traverse through a dynamical lens resulting in different phase responses. Reconstructing the lens becomes difficult in this scenario. For this reason, we consider the ideal case to be when the proper motion is less than the time span over which baseband data is recorded by an instrument. For CHIME this is approximately $\sim 20$ s \cite{Michilli2021}. Suppose the coherence scale is set by the diffractive scale of scintillation on the order of $\sim 10^8 $ cm and we care about the decoherence of an FRB which will be at an effective angular diameter distance of 1 kpc for the source located 1 Gpc in comoving distance from the observer, with a lens screen 1 kpc from the observer. In this example, the proper motion must be less than,
\begin{equation}
\dot{\theta} <  10^{-10} ~\mathrm{arcsecond/ s} \bigg(\frac{r_{coh}}{10^8 ~\mathrm{cm}}\bigg) \bigg(\frac{D_{eff}}{1 ~\mathrm{kpc}}\bigg)^{-1} \bigg(\frac{\Delta t}{20 ~\mathrm{s}}\bigg)^{-1} .
\end{equation}
Note this is for the maximum observable timescale. When lensing creates a path delay between images, the relevant timescale is the largest delay in arrival time between images, which will be on timescales of $\sim 10$ milliseconds at most in this paper. This will relax the proper motion condition to $\dot{\theta} <  10^{-7} ~\mathrm{arcsecond/ s}$. This implies a transverse velocity at the effective distance of $\sim 10^{4} ~\mathrm{km / s}$  for a coherence scale of $10^8$ cm. Given a small coherence scale ($\sim 10^6 ~\mathrm{cm}$) then the proper motion does matter. However, we will assume we are in a regime where it does not matter for the rest of this work. 

Observations of some pulsars have shown that the decoherence time of the transfer function of the scattering screen is on the order of minutes \cite{Main2017}. If proper motions do matter and the coherence of the propagation is lost, a time-lag correlation will not provide enough information about the lensing system. In these cases, using the secondary spectra \cite{Walker2004} is more ideal. Compared to pulsars, FRBs have a lower duty cycle such that this method is more difficult but can be done, e.g. as shown by \textcite{Wu2024}.
 
These considerations detail the spatial decoherence effects. In the time domain, we assume there is no expectation for the phase of the unlensed electric field to correlate in time, $\left <E(\bm{x}'^S,t)E^*(\bm{x}'^S,t-\hat{t}) \right > = 0 ~, \hat{t} \neq 0$ such that the phase correlations must come from the phase delay from the path difference of images. The expectation for the correlation for a single frequency channel is given by,
\begin{equation}\label{eq:expectcorr}
\begin{split}
    &\left < C(f_k,\hat{t})\right > = \int dt \left <V(f_k,t) V^*(f_k,t-\hat{t})\right > .\\
\end{split}
\end{equation}
Recalling equation \ref{eq:baseband} , the expected time-lag correlation is given as,
\begin{equation}\label{eq:corrfull}
\begin{split}
    &\left <C(f_k,\hat{t})\right > =  \left <N(f_k,t)N^*(f_k,t-\hat{t})\right > \\
           &+ \int df \left < |S(f)|^2 \right > \left < |H(f)|^2 \right > |P(f- f_k )|^2 e^{-\i 2 \pi f \hat{t}}, \\
\end{split}
\end{equation} 
where the first term is the expected noise correlations in the time-lag domain. The second term, which is often not accounted for in lensing studies, is the continuous representation of the signal in the time-lag domain for a frequency channel. With an instrument channel response, it is possible to have phase correlations in time due to spectral leakage, e.g. for CHIME the polyphase filterbank (PFB) introduces time-lag correlations at $\pm 2.56 ~\mathrm{\mu s}$ and $\pm 5.12 ~\mathrm{\mu s}$ \cite{Bandura2016,Kader2022}. We assume that the unlensed signal does not correlate in time, therefore the remaining phase correlations in the time-lag domain arise from the $\left < |H(f)|^2\right >$ term. With equation \ref{eq:transferfunc}, the expectation of the transfer function is,
\begin{equation}\label{eq:expectedtf}
\begin{split}
\left < |H(f)|^2\right > = \left <\sum_l \sum_m \bar{\varepsilon}^l \bar{\varepsilon}^{*,m} e^{\i 2 \pi f (\bar{\tau}^l -\bar{\tau}^m)} \right > .\\
\end{split}
\end{equation} 
We required that the index of refraction be less than unity and that the medium of lensing be smooth. These conditions imply that the lensing medium must have some mean change to the index of refraction $\bar{n} = \left < n \right >$, and some fluctuations described by a random variable, $\chi_n$,  $ n = \left < n \right > + \chi_n $, where $\frac{\chi_n}{\bar{n}} \ll 1$ \cite{Lee1975}. For lensing through a single stochastic medium satisfying these conditions, the expectation of the total magnification must be of order unity, $\left < \sum\limits_{l,m} \bar{\varepsilon}^l \bar{\varepsilon}^{*,m} \right > \sim 1$, as the total flux from propagation must be preserved. For the phase, the transfer function is considered incoherent if $\left <\sum\limits_{l,m} e^{\i 2 \pi f (\bar{\tau}^l -\bar{\tau}^m)}\right > \sim 0$ and coherent if all phase relations are preserved, $\left <\sum\limits_{l,m} e^{\i 2 \pi f (\bar{\tau}^l -\bar{\tau}^m)}\right > = \sum\limits_{l,m} e^{\i 2 \pi f (\bar{\tau}^l -\bar{\tau}^m)}$.

For an observation at a frequency $f_k$, the expectation for the phase coherence relies on the channel bandwidth, $f_W$, associated with the instrument channel response, $P(f-f_k)$. The channel bandwidth is much less than the observation channel frequency, $\frac{f_W}{f_k} \ll 1$. Within the channel bandwidth, for a coherent phase correlation to be measured, we require that the image paths are stationary over the channel bandwidth, $\gamma^l(f+f_k) \approx \gamma^l(f_k), |f- f_k | < \frac{f_W}{2}$. This implies the magnification for a path is $\bar{\varepsilon}^{l}(f+f_k) \approx \bar{\varepsilon}^{l}(f_k)$ and the delay is $\bar{\tau}^{l}(f+f_k) \approx \bar{\tau}^{l}(f_k)$. Then within a channel bandwidth, the phase relation is constant such that $
\left < |H(f)|^2\right > = \sum\limits_{l,m} \bar{\varepsilon}^l \bar{\varepsilon}^{*,m} e^{\i 2 \pi f (\bar{\tau}^l -\bar{\tau}^m)} $. The general requirement for phase coherent lensing observations can then be stated as the phase relation must be preserved on scales larger or equal to the frequency channel bandwidth. Given this context, narrow channel bandwidths and sharp channel responses are preferred for lensing phase searches to mitigate phase decoherence.

\section{Implementation} \label{sec:code}
We detail how the algorithm works in this section. The code is written in C++ but provided as a Python package through Cython \footnote{The source code is available at \url{https://github.com/zkader/RWLensPy} or installed directly through PyPI}. We outline the general algorithm below where in section \ref{sec:theory}, we have detailed the formalism used in the simulation code:

\begin{enumerate}
    \item We create an $N_x\times N_x$ grid to parameterize the multiple lens planes, $\bm{x}^i$. For each plane\footnote{Note that by choosing a characteristic achromatic lens scale for each plane, $\eta_i$, all planes can use the same $\bm{x}^i$ grid.} we have an achromatic lensing function, $\Phi_i(\bm{x}^i)$. The array is made to be contiguous. 
    \item We create another array of $N_f$ frequencies to be evaluated. For computation, we run simulations in parallel over the frequency array.
    \item We compute the gradient for each of the lens functions $\partial^i_a \Phi_i$ and the quantities $\partial^i_{11}\Phi_i +\partial^i_{22}\Phi_i \pm \sqrt{(\partial^i_{11}\Phi_i- \partial^i_{22}\Phi_i)^2 + 4 (\partial^i_{12}\Phi_i)^2}$. These quantities are, respectively, needed to determine the stationary points and eigenvalues of the Hessian for each frequency \footnote{It should also be noted that these quantities are determined for all grid points and saved to memory, where the search grid will be $N_x -4 \times N_x -4$ due to second order finite difference being used to evaluate the Hessian elements, $\partial^i_{ab}\Phi_i$.}. 
    \item In parallel for each of the $N_f$ frequencies, we perform a grid search for the stationary points. The images are found as follows, starting from the last lens plane $N$ with a given source position $\bm{x}'^{S}$ and iterating to the first lens plane:
    \begin{enumerate}
        \item For plane $i$, there is a set of source points $\bm{x}'^{i+1}$. A stationary point is identified if the gradient of the Fermat potential contains a root within the neighborhood of the cell. This is done using equation \ref{eq:gradpnt} to find all roots, $\bm{x}'^{i}$, for plane $i$ given source position $\bm{x}'^{i+1}$.  
        \item For image point $\bm{x}'^{i}$, the path delay through the $i$-th lens plane is found with equation \ref{eq:multidelay} and the magnification is found using equation \ref{eq:magni}.
        \item For each stationary point, $\bm{x}'^{i}$, the search is repeated for the next plane, $i-1$, where the point $\bm{x}'^i$ is now treated as the ``source" position in equation \ref{eq:gradpnt} for the next plane. 
        \item The final image position for a single stationary path, $\gamma^l$, in angular position will be $\eta_1 \bm{x'}^1$. The final delay is  $\bar{\tau}^l=\sum_i^{N} \tau^l_i$ and the final magnification is $\bar{\varepsilon}^l=\prod_i^{N} \varepsilon_i^l$. This process is repeated for all stationary paths resulting in $N_l$ total images.
   \end{enumerate}    
    \item To obtain the total propagation transfer function, equation \ref{eq:transferfunc}, all stationary paths for a given frequency are summed to obtain a singular transfer function, $H(f)=\sum_l^{N_l}H(\gamma^l,f)$. This is done assuming the image separations are less than the beam width, i.e. all the images are spatially unresolved by the telescope. Separately, the simulation tool can provide the total set of stationary points and their final spatial positions, time delays, and complex magnifications for each frequency.
\end{enumerate}

With the propagation transfer function, $H(f)$, equation  \ref{eq:baseband} is used to obtain the baseband data. Our FRB is modeled as a Gaussian process in the voltage timestream where, $\left <V(t)\right > = 0$ and $\left <|V(t)|^2\right > = A \exp\left(-\frac{(t - t_S)^2}{\sigma_S^2} \right)$, where $A$ is the square root of the signal to noise ratio, $t_S$ is the unlensed arrival time, and $\sigma_S$ is the temporal intrinsic width of the burst. For the baseband examples presented in this work, we use the same simulation framework for CHIME's baseband and signal processing chain as reported in \citet{Kader2022}. As a quick summary, the CHIME baseband simulation takes a real-valued voltage timestream sampled at 1.25 ns and channelizes the timestream into 1024 frequency channels between 400 and 800 MHz with a complex-valued timestream now sampled at 2.56 $\mu$s.

In figures \ref{fig:sim_error} and \ref{fig:sim_error_freq}, we present the benchmarks of the simulation with realistic expectations. Further details of the tests can be found in the related sections. The simulation works well in determining delay values to, at most, a fractional error of $10^{-4}$ across all frequencies above grid sizes of $1000 \times 1000$. The magnification values have larger error values, being at worst, a 10\% deviation of the expected value above grids of size of $1000 \times 1000$. The errors decrease with increasing grid sizes but will plateau once simulation precision is reached, mainly dictated by gradient and Hessian estimation. 

These errors imply that the simulations are best used in a general sense, precise values shouldn't be extracted but general morphologies and correlation responses should hold. A simulation can be said to be precise, for a realistic telescope simulation, if the magnification error is less than the noise environment of a telescope and if the delay error is smaller than the voltage sampling rate of the telescope. This statement depends highly on the type of lens, the strength of the lens, and the spatial grid resolution, and can be evaluated for each case.

For precise modeling of the lensing function, the simulation would require fine grid resolutions and possibly higher order gradient and Hessian estimates. However, when considering real data and realistic lensing scenarios, the noise contribution will affect magnification errors more than numerical precision. In general, this simulation can capture general morphological features and correlation signatures relatively quickly at the cost of numerical precision. Furthermore, for any lens case that is analytically known, one could substitute the analytic computation in a multi-lens scenario to further increase simulation accuracy.

\section{Analytic Test Cases}\label{sec:single}
Realistic lensing systems for FRBs will likely consist of multiple lensing components. For example, an FRB will have both a host and Milky Way scattering screen. Before we consider these complicated systems that are difficult or impossible to solve analytically, in this section we'll consider simple single-plane lensing cases with known solutions. We'll consider gravitational lensing from a point mass to highlight the achromatic morphology and frequency-independent phase correlation response as well as show benchmark tests for the simulation. We'll also consider a Gaussian plasma lens to showcase the chromatic images as they appear in baseband data and the chromatic phase correlation response. 

\subsection{Point Mass Gravitational Lensing}
Gravitational lensing comes from the warping of space-time causing deflections in the path of propagation. For thin plane lensing, the metric,
\begin{equation}
    ds^2 = -\left(1 + \frac{ 4 G M }{ c^2 r}  \right) c^2 dt^2 + \left(1 + \frac{ 4 G M }{ c^2 r} \right)^{-1} dl^2 ,
\end{equation}
has a weak field perturbation such that we can obtain the Fermat potential for a point mass lens as \cite{Schneider1992}, 
\begin{equation}
    T = \mu_g \frac{1}{2}(\bm{x}^1 -\bm{x}^S)^2 - \mu_l ln|\bm{x}^1| ~.
\end{equation}
We have defined $\eta^2 = \theta_E^2 =  \frac{4 G M D_{1S} }{c^2  D_{OS} D_{O1} } $ so that $\mu_g = \frac{(1+z_l) D_{OS} D_{O1} \theta_E^2}{c D_{1S}}$,  $\mu_l = \frac{4 G M (1+z_l)}{c^3}$, and $\alpha = 1$. The point mass system can be solved analytically. We can obtain the stationary condition from the lensing equations in section \ref{sec:theory},
\begin{equation}
    x^{s}_a = x^1_a - \frac{x^1_a}{|\bm{x}^1|} ~,
\end{equation}
where the solutions to this equation are the spatial image positions along the axis of the images,
\begin{equation}
    x'^{1,(+,-)} = \frac{|x^S|}{2} \pm \sqrt{(|x^{S}|)^{2} -4 } ~.
\end{equation}
The Hessian eigenvalues (eq. \ref{eq:eigen}) are 
\begin{equation}
    \lambda_{\pm} = 1 \pm \frac{1}{|x^1|^2} ~,
\end{equation}
where the magnification for any stationary path is  $\varepsilon = \frac{1}{\sqrt{\lambda_{+} \lambda_{-}}}$.

For our test case, let us construct a lensing system with a 10 solar mass lens located at redshift $z=0.42761$, half the comoving distance to a source at $z=1$ making $D_{\mathrm{eff}} = 6791 ~\mathrm{Mpc}$. The FRB point source sits at a unitless angular position of $x^S=2.5$.

\begin{figure}[!htb]
    \centering
    \includegraphics[width=\columnwidth]{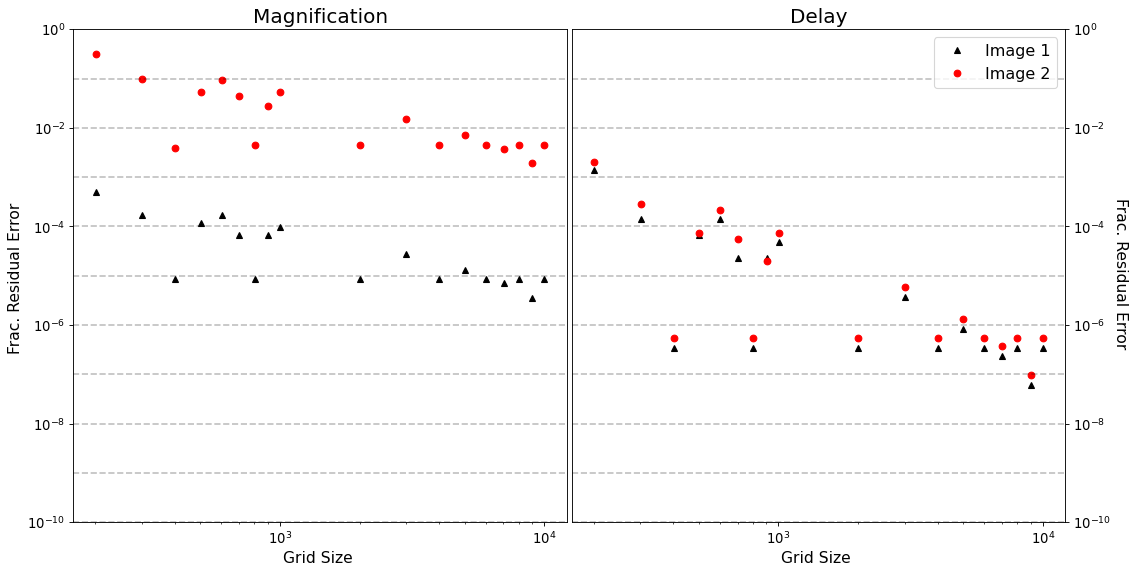}
    \caption{ Fractional error in the magnification, $\varepsilon^l$ (left), and delay, $\tau^l$ (right), of the two images (red circle and black triangle) compared to the analytic expectation. The fractional error reduces with increasing grid size but begins to plateau due to the numerical precision around a grid size of $\sim 2000 \times 2000$. 
    \label{fig:sim_error}}
\end{figure}

In figure \ref{fig:sim_error}, we show the fractional error for the two images, marked with red circles and black triangles, between the analytic values and the simulated values as a function of grid size. The magnification fractional error is shown in the left panel and the delay fractional error in the right panel. We find the fractional errors, for a grid size of $1001 \times 1001$, to be on the order of $10^{-3}$ for the magnification fractional error and $10^{-5}$ for the fractional time delay error. Degrading the grid size increases the error until $101 \times 101$ where the grid resolution was too coarse to find the two images. Increasing the grid size to $2001 \times 2001$, we see the fractional errors begin to plateau in value due to the numerical precision of the simulation.

\begin{figure}[!htb]
    \centering    
    \includegraphics[width=\columnwidth]{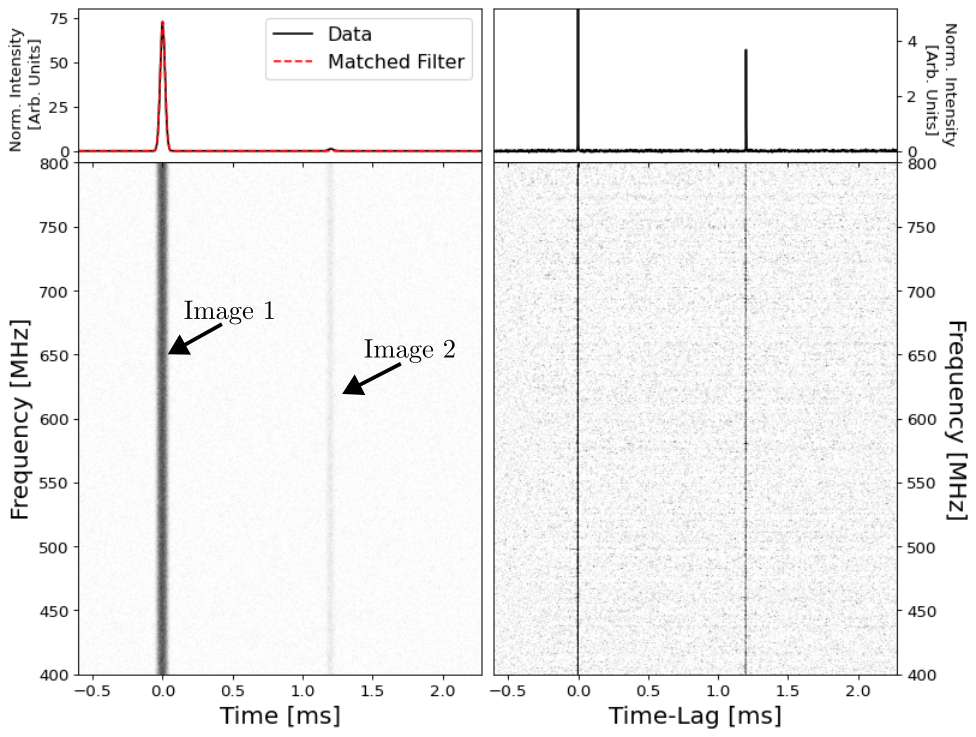}
    \caption{ Simulated waterfall (left) and time-lag correlation (right) of lensing from a point mass gravitational lens. The point mass lens produces two achromatic images where the delayed image is less magnified as compared to the primary image in the left panel. A search for phase correlations creates a perfect delta response at the associated delay between images, as seen in the time-lag correlation plot on the right. In the top panels, the intensity is normalized to the respective noise in the domain.}
    \label{fig:pm_dual}
\end{figure}

With an understanding of our simulation error, we can now evaluate the morphologies and correlation expectations of lensing scenarios. The transfer function for this lens is,
\begin{equation}\label{eq:gravtf}
    H(f) = \varepsilon^{(+)} e^{\i 2 \pi f \tau^{(+)}} + \varepsilon^{(-)} e^{\i 2 \pi f \tau^{(-)}} ~.
\end{equation}
From this, we can construct the baseband data using equation \ref{eq:baseband} as 
\begin{equation}\label{eq:basegrav}
\begin{split}
        V(f_k,t) &= \varepsilon^{(+)} \int df S(f) P(f-f_k) e^{-\i 2 \pi f (t-\tau^{(+)})} \\ 
        &+\varepsilon^{(-)} \int df S(f) P(f-f_k) e^{-\i 2 \pi f (t-\tau^{(-)})} \\ 
        &+  N(f_k,t) .\\
\end{split}
\end{equation}
From this equation, shown in the left panel of figure \ref{fig:pm_dual}, the original signal is altered such that there are two images, at $0$ and $\sim 1.2 $ ms. The delayed image is less magnified from the lens and therefore appears dimmer. As gravitational lensing is achromatic by nature, the delay and magnification are constant across frequency.

From the baseband data (eq. \ref{eq:basegrav}), suppose we window around the brighter burst, explicitly isolating the first term in equation \ref{eq:basegrav} and then search the time-lag correlated data (right panel of figure \ref{fig:pm_dual}). With equation \ref{eq:corr}, we have the expectation for the correlation,
\begin{equation}
\begin{split}
    &\left <C(f_k,\hat{t})\right > = |\varepsilon^{(+)}|^2 \int df \left <|S(f)|^2\right > |P(f-f_k)|^2 e^{-\i2\pi f \hat{t}}  \\
    &+ \varepsilon^{(+)}\varepsilon^{*(-)} \int df \left <|S(f)|^2\right > |P(f-f_k)|^2  e^{-\i2\pi f (\hat{t} - \Delta\tau )}  \\
    &+ \int dt \left <N(f_k,t) N^*(f_k,t-\hat{t})\right > ,\\
\end{split}
\end{equation}
where $\Delta\tau = \tau^{(-)} - \tau^{(+)}$ and the signal is assumed to be uncorrelated with the noise. The first and third terms will only contribute to the zero-lag correlation value, assuming the noise does not correlate in time. The second term is a temporal shift of $\Delta\tau$ in the Fourier frequency domain. This delayed response is exactly the signal auto-correlation response where gravitational lensing has only modified it up to a complex scaling factor. This is shown in the right panel of figure \ref{fig:pm_dual} as the delta-like response across all frequency channels at $\sim 1.2$ ms. The phase of this correlation also contains information about the Fermat potential in a topological manner \cite{Nakamura1999}. In this case $\varepsilon^{(+)}\varepsilon^{*(-)} = |\varepsilon^{(+)}||\varepsilon^{*(-)}|e^{\i\frac{\pi}{2}}$, which is the Morse phase difference between a local minimum and a saddle point. In summary, gravitational lensing provides a unique signature that should be identifiable directly from the phase correlation response in the time-lag domain.

\subsection{Gaussian/Rational Refractive Plasma Lens}
The Gaussian refractive plasma lens \cite{Clegg1998} is an empirically motivated model that has been useful in representing the extreme scattering events seen in quasars \cite{Fiedler1987} and pulsars \cite{GrahamSmith2011}. Plasma lenses arise from inhomogeneities in the electron density causing changes in the phase velocity of propagating waves, which corresponds to a non-unity index of refraction. As plasma lenses do not obey a Poisson equation like gravitational lenses there is no point mass equivalent. The closest equivalent is a Gaussian distribution, which is effectively a delta function in the limit the Gaussian width goes to zero. This lens is sometimes approximated as a rational lens \cite{Jow2023,Feldbrugge2023}.

Plasma lensing is sourced by free electrons. The plasma itself is assumed to be cold such that the plasma frequency is $\omega_p = \sqrt{4 \pi r_e c^2 n_e} $ (in SI units), where $r_e$ is the classical electron radius and $n_e$ is the electron number density. For dispersion due to this cold plasma, we look at the group velocity
\begin{equation}
    v_g \equiv \frac{\partial \omega}{\partial k} = c \left(1 - \frac{\omega_p^2}{\omega^2}\right)^{1/2} ~.
\end{equation}
Using this, the Fermat potential for the Gaussian distributed plasma lens is\cite{Clegg1998},
\begin{equation*}
    T = \mu_g \frac{1}{2}(x^1_a -x^S_a)^2 + \mu_l \exp\left( -\frac{x_a^{1}}{2 \sigma^2} \right) ~.
\end{equation*}
For the rest of this section, we define $\sigma = 1$ such that it is the angular size of the lens. Then we set $\mu_g = \frac{(1 + z_l)D \theta_a^2}{c}$,  $\mu_l = \frac{(1 + z_l)k_{\mathrm{DM}} }{ f^2} \mathrm{DM_0} $, and $\alpha =  \frac{k_{\mathrm{DM}} \mathrm{DM_0} c }{ D \theta_a^2 f^2}$. We additionally note the dispersion measure is $\mathrm{DM} = \int n_e  dl $.

Unlike the point mass gravitational lens, there is no analytic solution to obtain the spatial position of images, except for $y=0$. However, $y=0$ is a critical degenerate point and we would need full wave optics considerations to get the propagated electric field. Instead, we can make use of the rational lens for our test case which does have an analytic solution. The Fermat potential is given as \cite{Jow2023,Feldbrugge2019}, 
\begin{equation}
    T = \mu_g \frac{1}{2}(x^1_a -x^S_a)^2 + \mu_l \frac{1 }{1 + \frac{(x^1_a)^2}{ 2} }  ~,
\end{equation}
where the parameter definitions are the same as before. The lens equation is given as
\begin{equation}\label{eq:rationalimgs}
    x^S_a = x^1_a - \alpha \frac{x^1_a}{\left( 1 + \frac{(x^1_a)^2}{2} \right)^{2} }
\end{equation}
which means, along the axis of the source point, images are found at the roots of the polynomial of the fifth order,
\begin{equation}
    0 = \frac{1}{4 }x^5 - \frac{x^S}{4 }x^4 + x^3 - x^S x^2 +\left( 1 - \alpha \right) x - x^S .
\end{equation}
The real solutions to this polynomial are the image positions that can be found using polynomial root-finding algorithms. Similar to the previous section, the magnification is given by $\varepsilon = \frac{1}{\sqrt{ \lambda_+ \lambda_- }} $, where the Hessian eigenvalues are,
\begin{equation}
    \lambda_{\pm} = 1 + \frac{\alpha}{(1+\frac{1}{2} x^2)^2 }\left( \frac{x^2}{(1+\frac{1}{2} x^2)}(1\pm1) - 1 \right) .
\end{equation}

\begin{figure}[!htb]
    \centering
    \includegraphics[width=\columnwidth]{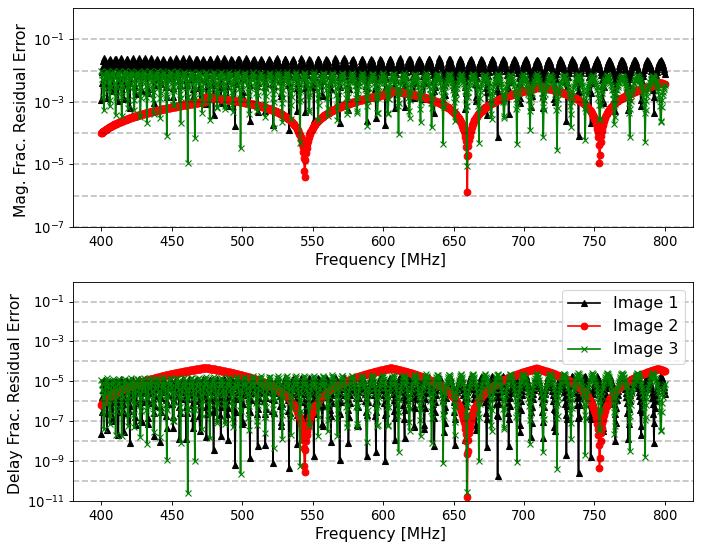}
    \caption{ Fractional error in the magnification, $\varepsilon^l$ (left), and delay, $\tau^l$ (right), of the three images (colored shapes). On a grid ($1001 \times 1001$) the computation error remains relatively consistent across frequency. The maximal image error values are consistent with the gravitational lensing example (Fig. \ref{fig:sim_error}). }
    \label{fig:sim_error_freq}
\end{figure}

We construct our lensing system by having the FRB at a redshift of 1 and the lens a kpc from the FRB. We set $\theta_a = \frac{10 ~\mathrm{AU}}{D_{ol}}$, $y=1.5$, $\sigma=1$ and $\mathrm{DM_0}=0.45 ~\mathrm{pc} ~\mathrm{cm}^{-3} $. Using these parameters and our derived equations, we can compare the fractional error in parameter recovery, which is shown in figure \ref{fig:sim_error_freq}. This system produces three images and from this, we find the fractional errors, at worst, agree with the order found using the same grid size for the gravitational lens, figure \ref{fig:sim_error}, and the errors are consistent across frequency. The main conclusion of this test showcases that the relative phase errors across frequency are consistent with the numerical error associated with the grid resolution and numerical precision.

\begin{figure}[!htb]
    \centering    
    \includegraphics[width=\columnwidth]{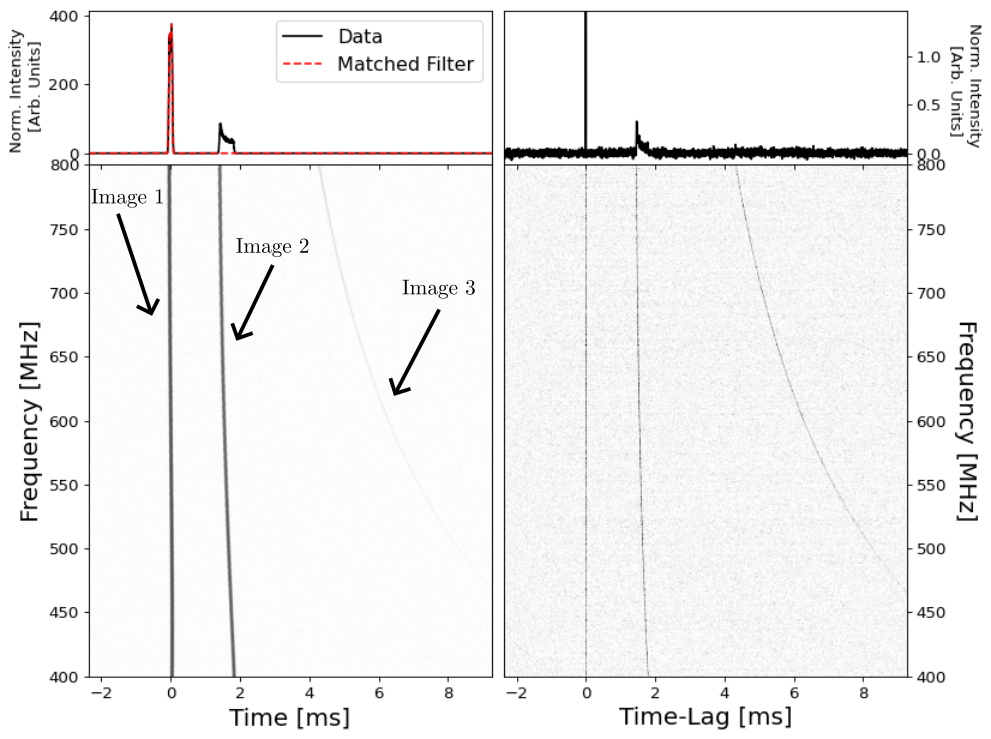}
    \caption{ Simulated waterfall (left) and time-lag correlation (right) of lensing from a refractive rational lens. This lens configuration produced three images where the two delayed images are successively less magnified, as seen in the left panel. Phase correlations, shown in the right panel, show images are correlating in time with a dispersive frequency relation. The magnification also changes as a function of frequency where the last image can be visibly seen to get dimmer with frequency.} 
    \label{fig:plasma_dual}
\end{figure}

This test case features images with a frequency-dependent transfer function such that we can use it to highlight effects due to chromatic propagation. The general transfer function can be written as,
\begin{equation}\label{eq:plasmtf}
    H(f) = \sum_l \varepsilon^{l}(f) e^{\i 2 \pi f \tau^{l}(f)} ,
\end{equation}
where the key difference in comparison to the gravitational lensing transfer function, equation \ref{eq:gravtf}, is the magnification and delays of images are frequency dependent. It should be noted that the number of images is also frequency dependent as seen in equation \ref{eq:rationalimgs} where $\alpha$ is dependent on frequency. The implication of this is that by observing over a large enough frequency range, one might see the merging of images which would correspond to a caustic point. Our simulation works in the Eikonal limit so we will not consider modeling this behavior in this work. For baseband data, we use equations \ref{eq:baseband} and \ref{eq:plasmtf}  to obtain,
\begin{equation}\label{eq:basebandplasm}
\begin{split}
        &V(f_k,t) = N(f_k,t)\\
        &+ \sum_l \int df \varepsilon^l S(f) P(f-f_k) e^{-\i 2 \pi f (t-\tau^{l})}.\\
\end{split}
\end{equation}
If the image point, $x'^1$, does not shift sufficiently as a function of frequency then the response will resemble a dispersion sweep and follow a dispersive time relationship, where the delay is scaling with $f^{-2}$. Conversely, if the spatial position does shift significantly, this might not hold. This occurs because the magnifications and delays are dependent on the image positions as a function of frequency. This has direct implications to the response in the time-lag domain where the total correlation response at a single time-lag will be the chromatic response between all pairs of images. In figure \ref{fig:plasma_dual}, image 1 is the burst at $0$ ms while images 2 and 3 are bursts seen to be dispersed from $3$ and $5$ ms, respectively. Image 3 is seen to be the faintest and most dispersed image. We select the bright burst with the earliest arrival time and label it as image $1$. Using equation \ref{eq:basebandplasm}, the expectation for the time-lag correlation is,
\begin{equation}\label{eq:corrplasm}
\begin{split}
    &\left <C(f_k,\hat{t})\right > = \int df |\varepsilon^{1}|^2 \left <|S(f)|^2\right > |P(f-f_k)|^2 e^{-\i2\pi f \hat{t}}  \\
    &+ \sum_{l=2}^3 \int df \varepsilon^{1}\varepsilon^{l,*} \left <|S(f)|^2\right > |P(f-f_k)|^2  e^{-\i2\pi f (\hat{t} - \Delta\tau^{l1})}  \\
    &+ \int dt \left <N(f_k,t) N^*(f_k,t-\hat{t})\right > ,\\
\end{split}
\end{equation}
where the first term is the zero-lag auto-correlation, the second summation term is the correlation of the selected image with the other images, and the third term is the noise auto-correlation. The time-lag correlation can be seen in the right panel of figure \ref{fig:plasma_dual}. The lens strength is proportional to the frequency of the channel center, $\alpha \propto f_k^{-2}$. The channel bandwidth is significantly smaller than the channel center, $\frac{f_W}{f_k} \ll 1$. Therefore, we can assume images are stationary in frequency across a single channel, $x'^{l}_a(f) \approx x'^{l}_a(f_k)$ and the magnification is constant, $\varepsilon^{l}(f) \approx \varepsilon^{l}(f_k)$. At points of stationary phase, the difference between two image delays is
\begin{equation*}
\begin{split}
\Delta\tau^{l1}(f_k) &= \frac{\mu_l^2}{\mu_g} \left( ( \partial_a DM(x'^{l}_a) )^2 -( \partial_a DM(x'^{1}_a) )^2 \right)\\
&+ \mu_l \left( DM(x'^{l}_a)- DM(x'^{1}_a) \right) .\\    
\end{split}
\end{equation*}
Assuming the path delay between images is dominated by the second term, which has a frequency scale of $f^{-2}$, the phase response across the time-lag correlation is a dispersive phase response, i.e. it follows a dispersion sweep with some effective DM ($t \propto k_{\mathrm{DM}} \mathrm{DM}_{\mathrm{eff}} f^{-2}$).

For the time-lag correlation in the right panel of figure \ref{fig:plasma_dual} we can see the two delayed images correlating with the initial bright burst. The offset from zero-lag corresponds to the spatial offset between the images and the dispersive sweep across the frequency corresponds to the effective differential DM between images. This provides a method to reconstruct the Fermat potential and therefore model the lens by measuring the effective DM values between images and their geometric delay. For example, one could use dedispersion algorithms to find the differential DM between images and then find the geometric delay and along with the magnification use these measurements to recover lens parameters. 

This also has implications for a gravitational lensing search using phase correlations. As gravitational lensing is achromatic by nature, it is effectively a geometric delay. If there existed an appreciable differential DM between the two image paths, then the gravitationally lensed FRB would appear as a plasma lensed FRB, except with only two images. However, if only two plasma lensed images are seen, that is to say, image 3 is below the noise floor of the instrument, then the two cases are degenerate in both baseband data and the time-lag domain. The degeneracy could be broken through spatially resolving image positions, plasma lensing would chromatically shift images while gravitational lensing would not. The gravitational lens or plasma lens parameters could be recovered by dedispersing the second image and then performing a time-lag auto-correlation, though there is a model degeneracy for which one needs to account. Nonetheless, this case shows coherent phase responses from propagation can be measured in the time-lag domain.

\section{Case Studies}\label{sec:multi}
\subsection{Scattering Screens}\label{sec:scatt}
For our first case, we investigate a scattering screen with our simulation. Scattering will be the largest source of decoherence for phase coherent astrophysical lensing searches. This makes modeling scattering one of the most important goals for this work. 

For the scattering of radio waves, we can extend the formalism from the case of a refractive plasma lens by placing a distribution of inhomogeneities in the electron density distribution for cold plasma along the line of sight. If the scale of the inhomogeneities is small ( $\sim$AU) compared to the distance between the source and the observer then one can construct a scattering screen to represent the phase fluctuations imposed on the wavefront due to the electron density distribution \cite{Salpeter1967,Rickett1977,Narayan1992}. The Fermat potential for a scattering screen is similar to the refractive plasma lens, with the main difference being the electron densities are no longer uniform or analytic and are now described by a random distribution. The distribution itself can be characterized by various models. For example, one could assume some power law spectrum to generate the electron densities \cite{Narayan1992,Rickett1977} or a Gaussian distribution \cite{Salpeter1967}. It should be noted that observations of scattering screens from some pulsars do suggest a thin and highly anisotropic electron density spectrum, implying that a 1D screen can be used to model the phase fluctuations \cite{Walker2008,Brisken2010,Sprenger2022}. For this case, we choose to generate a 2D phase screen using Gaussian statistics.

We can create the complete DM contribution for a scattering screen as in \cite{Lee1975}, where our lensing potential is of the form,
\begin{equation}
    \Phi(\bm{x}^1) = \mathrm{DM}(\bm{x}^1) = \overline{\mathrm{DM}} + \delta\mathrm{DM}(\bm{x}^1) ,
\end{equation}
where $\overline{\mathrm{DM}}$ is the spatially independent average DM contribution (assuming a constant electron density distribution throughout the universe) and $\delta\mathrm{DM}(\bm{x}^1)$ is a stochastic, spatially varying distribution of DM values. The dispersive phase delay from the constant $\overline{\mathrm{DM}}$ is accounted for with dedispersive algorithms and would be removed for phase correlations such that we can ignore this contribution. What remains for the Fermat potential are the inhomogeneous fluctuations sourcing the scattering. 

We can define the Fermat potential for a scattering screen as,
\begin{equation}
    T = \mu_g \frac{1}{2}(x^1_a - x^S_a)^2 + \mu_l \delta\mathrm{DM}( x^1 ).
\end{equation}
Let us consider the scattering screen to have DM fluctuations that are Gaussian \cite{Lee1975}. In this case, $\left <\delta\mathrm{DM}\right > = 0$ and $\left <\delta\mathrm{DM}^2\right > = \sigma_{\mathrm{DM}}^2$. We place the source at a redshift of 1 and the screen 1 kpc from the source. The geometric parameter is given by $\mu_g = \frac{(1+z_l)D \theta_a^2}{c}$ and with a zero-mean, unity variance Gaussian random variable, we define $\mu_l = \frac{(1+z_l)k_{\mathrm{DM}} \sigma_{\mathrm{DM}} }{ f^2} $. Then, the lens strength is $\alpha = \frac{k_{\mathrm{DM}} c \sigma_{\mathrm{DM}} }{ D \theta_a^2 f^2}$. We can couple our desired scattering time at 400 MHz, $\tau_{\mathrm{scatt}}$, to the angular scale as $\theta_a = \sqrt{ c \tau_{\mathrm{scatt}} / D } $. Setting $\alpha = 1$, we set the variance of the DM fluctuations to be $\sigma_{\mathrm{DM}} = \tau_{\mathrm{scatt}} f_{\mathrm{ref}}^2 / k_{\mathrm{DM}} $.

We have a grid of $2001 \times 2001$ with a max scale of $ 5 \theta_a$. Note that the simulation assumes the Fermat potential below the resolution scale is a smoothly varying function. The DM fluctuations are set by a $\tau_{\mathrm{scatt}} = 0.1$ms at 400 MHz. Computationally, this is done by having each pixel on the lens plane be a random variable sourced from this Gaussian probability distribution. Scattering is typically expected to create a pulse profile that is exponential in time, $|E(t)|^2 \propto \exp(-\frac{t}{\tau_{\mathrm{scatt}}})$ \cite{Williamson1975}, where $\tau_{\mathrm{scatt}}$ is the characteristic scattering time.

\begin{figure}[!htb]
    \centering    
    \includegraphics[width=\columnwidth]{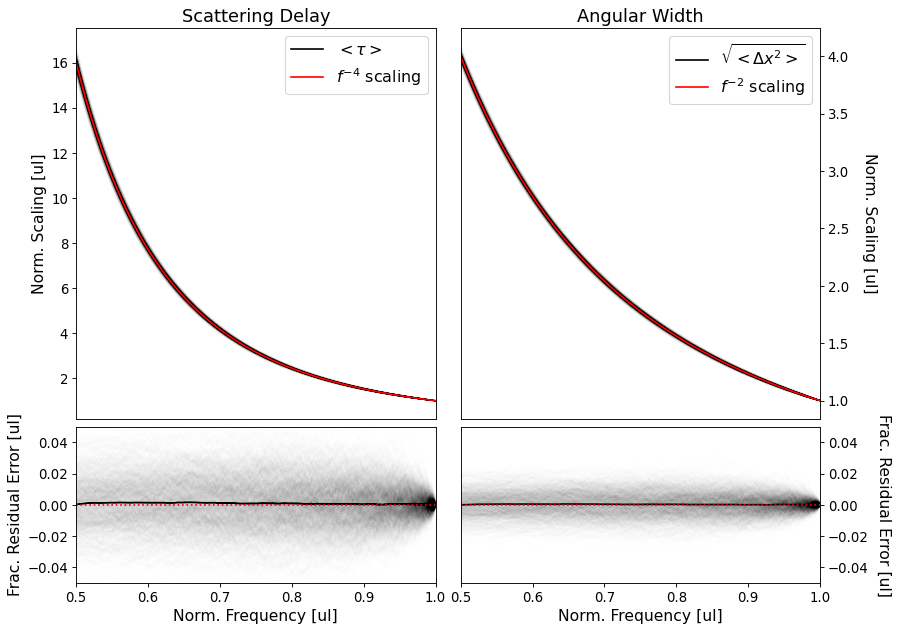}
    \caption{ In the left panels, the distribution (blue line) of the expected delay for images (black line) as a function of frequency is shown to, on average, follow the expected scattering relation for a Gaussian process, $\tau \propto f^{-4}$. Similarly in the right panels, the distribution (blue line) of the expected angular width of images (black line) as a function of frequency is shown to, on average, follow the expected scattering relation for a Gaussian process, $\sqrt{\left <\Delta x^2\right >} \propto f^{-2}$.  } 
    \label{fig:gscatt_scale}
\end{figure}

For plasma inhomogeneities generated from a Gaussian process, the scattering time scales in frequency as $\tau_{\mathrm{scatt}} \propto f^{-4}$  while the angular width associated with scattering scales as $\theta_{\mathrm{scatt}} \propto f^{-2}$ \cite{Salpeter1967,Rickett1977}. To test whether the simulation is following the expected frequency scaling relation, we create 1000 different realizations at 1024 frequencies. For each frequency, we can determine the mean image delay for each realization and normalize the scaling to the largest frequency. The resulting plot is shown in the left panel of figure \ref{fig:gscatt_scale}. The scaling for the image delay of all realizations is $\left <\tau\right > \propto f^{-4}$, as expected. The top right panel shows the angular width of the distribution where the frequency scaling also follows the expected scaling, $\sqrt{\left <\Delta x^2\right >} \propto f^{-2} $. Both bottom panels show the fractional difference between the expected scaling and the average scaling of 1000 realizations. The 1000 realizations are plotted along with the average of the realizations, shown in solid blue. 

\begin{figure}[!htb]
    \centering    
    \includegraphics[width=\columnwidth]{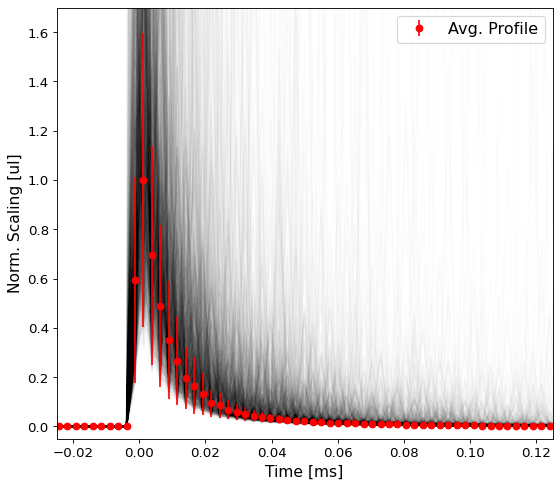}
    \caption{ The average pulse profile and standard deviation (red) are shown for the 1000 realizations (black) of a single scattering screen. The profile is sampled every 2.56 $\mu$s. At each time sample, there exists a probability for an image to form. The average profile follows an exponential but caustics and bright images are shown to vary the intensity by 50\%. } 
    \label{fig:gscatt_prof}
\end{figure}

In figure \ref{fig:gscatt_scale}, the simulation is shown to produce the expected frequency scalings and expected images from simulations of a stochastic distribution. In observations, these quantities aren't directly measured and instead, it is the propagation transfer function in the time domain $\left< |H(t)|^2 \right>$ (eq. \ref{eq:expectedtf}) that is observed. The pulse profile observed is, effectively, the image distribution of the transfer function weighted by the magnification, i.e. $\left< |H(t)|^2 \right> = \exp(-\frac{t}{\tau_{\mathrm{scatt}}}) $. In figure \ref{fig:gscatt_prof}, we generate $\left< |H(t)|^2 \right> $ for 1000 realizations to evaluate the range of profiles that could be generated, shown in black. The average profile and standard deviation are shown in red, for every 2.56 $\mu$s interval. The average profile follows the expectation but the intensity can vary by up to 50\%, depending on if an image formed near or at a caustic. One consequence of this is, with our formalism, the scattering timescale set for the simulation is the scale over which the majority of images are generated and not the characteristic $1/e$ timescale of the pulse profile. 

\begin{figure}[!htb]
    \centering    
    \includegraphics[width=\columnwidth]{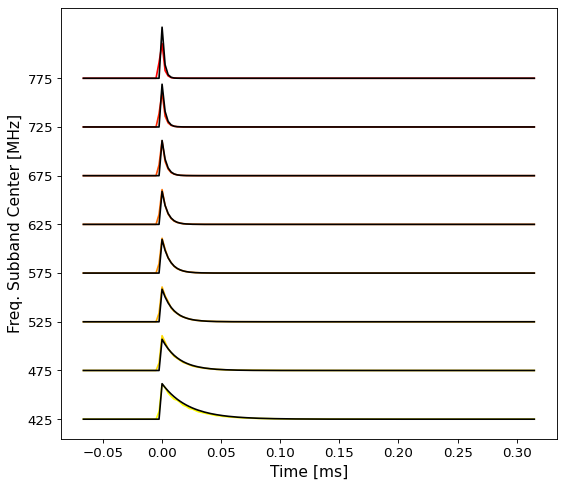}
    \caption{The fit (black) of an exponential pulse profile to the statistical average profile across 8 sub-bands (red to yellow). The fit finds a scattering time of $ 5.36 \pm 0.06 ~\mu$s at 600 MHz and a frequency scaling index $-3.99 \pm 0.05$.} 
    \label{fig:gscatt_scale_fit_imp}
\end{figure}

To measure the scattering time and frequency scaling, we average the 1000 realizations over 8 sub-bands across the 1024 frequency channels. Each frequency sub-band is 50 MHz in width. We fit an exponential simultaneously to all sub-bands. This is shown in figure \ref{fig:gscatt_scale_fit_imp} where the black lines are the fitted profile and the colored lines are the average profile in each sub-band. The statistical error is taken as the standard deviation and the fit has a reduced $\chi^2$ value of $0.1$. We obtain a scattering time of $ 5.36 \pm 0.06 ~\mu$s at 600 MHz and a frequency scaling of $-3.99 \pm 0.05$. For this case, the total timescale over which the majority of images are created was set to $19 ~\mu$s at 600 MHz, while the observed $1/e$ scattering time is smaller by, approximately, a factor of 4. The agreement of the scaling index follows the work of \textcite{Williamson1975}, where we have shown the simulations are statistically consistent in the Eikonal limit. 

\begin{figure}[!htb]
    \centering    
    \includegraphics[width=\columnwidth]{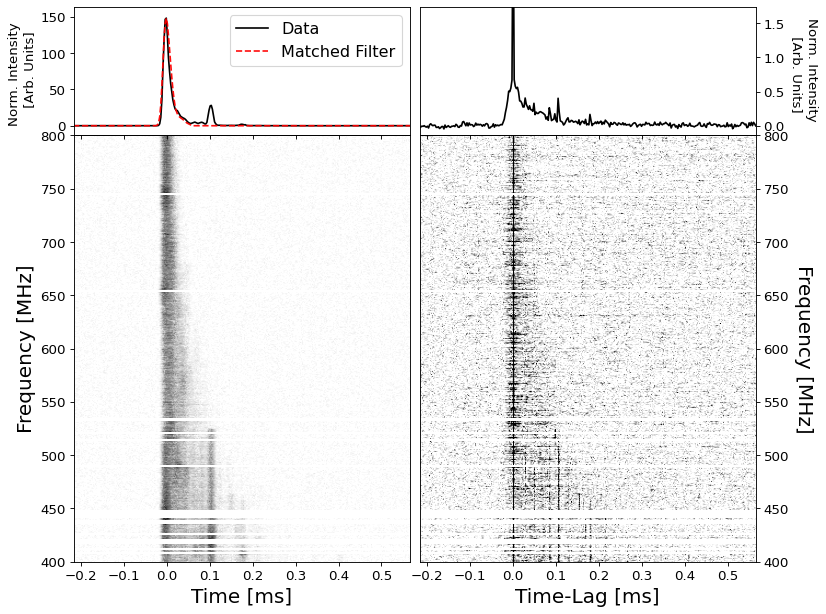}
    \caption{ Simulated waterfall (left) and time-lag correlation (right) of lensing from a scattering screen that is generated from Gaussian fluctuations. The screen produces multiple images such that the superposition of these images creates a tail with more images seen at lower frequencies. In the time-lag domain, there is excess variance but no distinct image. The discrete images are not correlating because no distinct phase relation is being preserved. Frequency channels with apparent caustics are masked. } 
    \label{fig:gscatt_dual}
\end{figure}

For a single realization and a more realistic dataset, we simulate the full transfer transfer function and baseband data. The morphology and time-lag correlation for this one realization is shown in figure \ref{fig:gscatt_dual} in the left and right panels, respectively. The goal with one realization is to test how consistent it is with the average expectation. In the baseband simulation, there are critical frequencies where the magnification asymptotes to infinity. We have flagged and masked these frequencies as they are caustics and regions where wave optics considerations would be required to predict the true magnification. The morphological structure of the scattered FRB is seen to have a scattering tail. The discrete sampling of the distribution generates resolved images, which can be seen in the left panel at $\sim$0.1 ms for the lower frequencies. 

\begin{figure}[!htb]
    \centering    
    \includegraphics[width=\columnwidth]{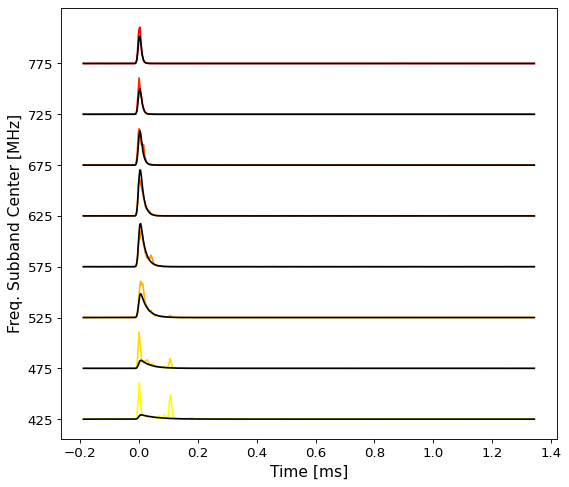}
    \caption{The fit (black) of an exponential pulse profile convolved with a Gaussian to the pulse profile across 8 sub-bands (red to yellow) for simulated baseband data. The fit finds a scattering time of $ 12.7 \pm 0.4 ~\mu$s at 600 MHz and a frequency scaling index $-4.2 \pm 0.1$.} 
    \label{fig:gscatt_scale_fit_ful}
\end{figure}

We can perform a morphological fit to obtain the scattering time and frequency scaling index as we did before. We find for the single scattering screen realization, a scattering time of $ 12.7 \pm 0.4 ~\mu$s at 600 MHz and a frequency scaling index $-4.2 \pm 0.1$. The fitted pulse profile for all sub-bands can be seen in figure \ref{fig:gscatt_scale_fit_ful}. The fitted profile (black) has a reduced $\chi^2$ of 0.1 but the fitted values are not consistent with the statistical expectation at the $1\sigma$ level. The discrepancy likely originates from the bright images, especially for the lower frequencies where the scattering power is spread across a larger timescale. In figure \ref{fig:gscatt_scale_fit_ful} the images are visibly present in the bottom two sub-bands centered at 475 and 425 MHz. The implication of this is our simulation can be discrepant for a single realization but scattering is statistically consistent, as shown in figure \ref{fig:gscatt_scale_fit_imp}.

For scattering simulations, there is another subtlety that one should take note of for grid discretization. In this simulation framework, there can only exist one image within a grid cell meaning everything below the grid resolution is treated as a point source. This has implications for a Gaussian process as there should exist a probability of creating an image at all spatial scales below the grid resolution. Morphologically, this would create a smoother distribution of images in time such that the lower frequency distribution in the left panel of figure \ref{fig:gscatt_dual} would resemble the distribution near 800 MHz. However, this is only true if the FRB is a true point source. An extended source will average the phase of the scattering screen over the size of the source creating an incoherent source. In other words, the simulation should be representative of the ensemble properties of scattering, provided the source size is smaller than the grid resolution. In this limit, the average phase over the source size is equivalent to the average phase of the grid cell, specifically the geometric term of the Fermat potential. 

In the time-lag correlation plot (right panel of figure \ref{fig:gscatt_dual}), the discrete images of the scattering screen can be seen to phase correlate. There is no clear and distinct image correlation unlike the point mass gravitational. Instead, there exists excess phase correlation power where one discrete image cannot be distinguished from another. The implication is that a static screen will produce phase correlation power within the scattering time. While phase correlations do exist the other important feature to note is the distribution of power. Due to energy conservation, the total magnification of all images will be unity on average. This implies the distribution of images over a larger scattering time will spread the power over that time, effectively making the FRB and the phase correlation signal dimmer. In summary, the simulation toolset can generate the expected morphologies from a scattering screen. In the time-lag domain, we show that the scattering screen generates no distinct coherent image unlike the point mass gravitational lens or the refractive Gaussian plasma lens. However, there does exist excess variance above the noise expectation in this domain.

\subsection{Host Scattering Screen and Milky Way Scattering Screen }
The morphology of FRBs contains information about both their host galaxy and the Milky Way as they propagate through the cold plasma in both environments. In this section, we will simulate a system to investigate this propagation. FRBs look to be associated with dense, magnetized regions such that in a two-screen model, the host screen contributes to the scattering while diffractive scintillation is caused by the Milky Way screen \cite{Masui2015}. 

With this prescription, we want to create a scenario where both diffractive scintillation and a scattering tail are observed. Instead of assuming a turbulent model, we'll create our DM fluctuations using Gaussian statistics. We generate a screen with a scattering time of $\sim 0.1$ ms at 400 MHz for the host screen to create a scattering tail and a scattering time of $\sim$10 nanoseconds for the Milky Way to create the diffractive scintillation. The physical system is set with the FRB source located at a comoving distance of 1 Gpc, the host screen 1 kpc from the FRB, and the Milky Way screen 1 kpc from the observer. We set our characteristic scale to be $\theta_a = \frac{0.1 \mathrm{AU}}{D_{eff,1}}$ for the MW screen and $\theta_a = \frac{50 ~\mathrm{pc}}{D_{eff,2}}$ for the host screen. This sets our DM fluctuations to be $1\times10^{-7} ~\mathrm{pc} ~\mathrm{cm}^{-3} $ for the MW screen and $5\times10^{-4} ~\mathrm{pc} ~\mathrm{cm}^{-3} $ for the host screen. The simulation is done on a grid of size of $301 \times 301$  with a resolution of $\frac{0.05 \mathrm{AU}}{D_{eff,1}}$ for the MW screen and $\frac{10 ~\mathrm{pc}}{D_{eff,2}}$. We assume that the Fermat potential is smoothly varying below this scale to allow for the computation of the gradient and Hessian.

\begin{figure}[!htb]
    \centering    
    \includegraphics[width=\columnwidth]{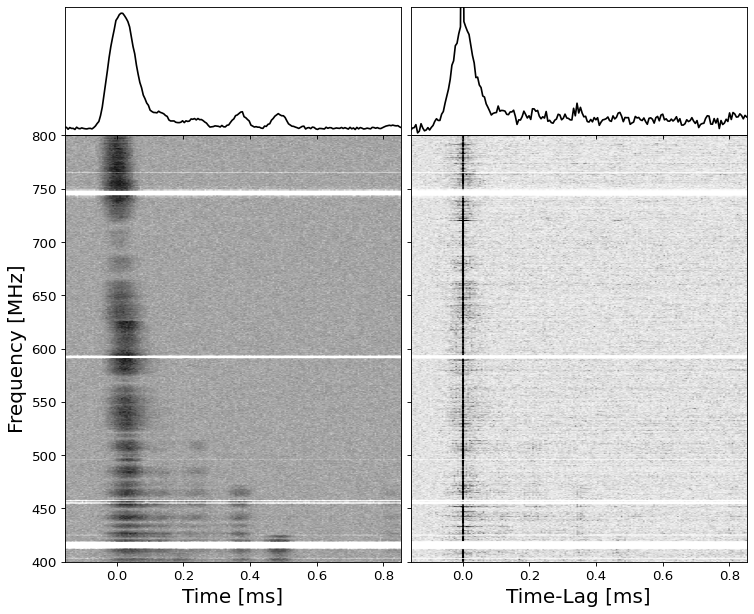}
    \caption{ Simulated waterfall (left) and time-lag correlation (right) of lensing from two scattering screens, both generated with Gaussian fluctuations, where one is placed at a kpc from the host and one at a kpc from the observer. The FRB is at a comoving distance of 1 Gpc. In this case, the host screen creates the scattering tail morphology at large timescales while the Milky Way screen produces the diffractive scintillation morphology. Phase correlations show some excess correlations being present, especially near the bottom of the band. Frequencies with caustics are masked. } 
    \label{fig:gscatt_gscatt_dual}
\end{figure}

The resulting propagation transfer function from simulated baseband data is shown in figure \ref{fig:gscatt_gscatt_dual}. For coherent multi-path scattering, the frequency scintillation scale, i.e. the size of the scintles and scale over which the brightness varies along the frequency spectra of the burst, is given by $\Delta f_{\mathrm{scintt}} \sim \frac{1}{2 \pi \tau_{\mathrm{scatt}}}$. This is seen in the left panel of figure \ref{fig:gscatt_gscatt_dual}. The Milky Way screen has $ \tau_{\mathrm{scatt}} < 2.56 \mu \mathrm{s}$, generating frequency modulations on the diffractive scintillation scale. The host screen has the larger scattering time ($\tau_{\mathrm{scatt}} > 2.56 \mu \mathrm{s}$), and is generating the morphological scattering tail.

The time-lag correlation, in the right panel of figure \ref{fig:gscatt_gscatt_dual}, shows there exist phase correlations between images within the time scale of the burst. For this system, the addition of the second screen has not removed the coherence of the other screen. We do note the total number of images being propagated for multi-plane scattering screens does require an increase in computational cost for efficient simulation time. The simulations for this case study, with a grid of $301 \times 301$ and $1024 \times 500$ frequency points, were done in $\sim 260$ core-hours. While the computational cost is high for generating complete baseband transfer functions, generating the morphology for only the center of frequency channels (1024 in this paper) is possible in $\sim 30$ core-minutes. We foresee an extension of this simulation to model and possibly fit for observed scattering morphologies for FRBs, especially those that exhibit both diffractive scintillation and a scattering tail. We leave for future work, the analysis of simulated two scattering screen systems, such as the analysis done for real systems in \textcite{Sammons2023}. A better understanding of these systems, especially whether phase coherence can be measured, can help constrain the FRB emission region \cite{Kumar2024}.

\subsection{Gravitational Point Mass and a Scattering Screen}
The gravitational lensing system of a point mass and a scattering screen is a crucial case for understanding how decoherence will impact a phase coherent search looking for gravitational lensed FRBs such as the search conducted in \textcite{Kader2022}. For coherent lensing, without an understanding of decoherence, we cannot quantify what has not been seen and so it is not possible to establish robust limits in the scenario of a non-detection. For this case study, we seek to investigate how the gravitational lensing phase correlation signal is altered with a scattering screen.

In \textcite{Leung2022}, the constraints on gravitational lenses larger than a solar mass suffer in sensitivity from the decoherence that would result from the presence of a scattering screen close to the FRB. There is a degeneracy in the case of a non-detection between whether there are no coherent lensing signals or whether the coherence of the lensing signal is erased by the presence of a scattering screen. The main concern for a phase coherent gravitational lensing search is whether coherence is maintained along the two paths of the gravitational lens. We want to explore the notion of whether the scattering screen is ``resolved" by the gravitational lens. That is to say, whether the angular extent of the scattering screen is comparable in scale to the projected Fresnel scale on the lens plane (see section \ref{sec:coherence}).

To demonstrate our simulation toolset's ability to model this, we will present three scenarios. To increase the computation speed and accuracy, we use the analytic equations for a gravitational lens to calculate its image positions, delay, and magnification of the images rather than a grid. In all scenarios, we will have the FRB at a comoving distance of 1 Gpc and a 10 solar mass gravitational lens placed at half the comoving distance from the FRB to the observer. We place the FRB source offset at $x^{S} = 1.5$ such that the delay between the gravitational images will be $\sim 0.7$ ms with an angular separation of $\sim 20$ microarcseconds. The scattering screen is sourced from Gaussian fluctuations of zero-mean as done in section \ref{sec:scatt}. Using the formalism from section \ref{sec:scatt}, we simulate several choices for the $\sigma_{\mathrm{DM}}$ parameter and distance to the scattering screen.

Scenario 1 is the case where both the angular extent of scattering is smaller than the angular separation of the gravitational lens images and the temporal broadening of scattering is smaller than the time delay between gravitational images. $\tau_{\mathrm{scatt}} = 0.1 \mu$s and the scattering screen half the comoving distance from the gravitational lens to the FRB. Scenario 2 is the case where the angular extent of scattering is comparable to the gravitational image separations and the scattering time is comparable to the gravitational time delay between images. The scattering screen is placed at half the comoving distance between the gravitational lens and the FRB and $\tau_{\mathrm{scatt}} = 1 $ms. Finally, in scenario 3, we have the $\tau_{\mathrm{scatt}} = 0.1 $ms as in scenario 2 but the scattering screen is placed 1 kpc from the FRB. In this case, the angular extent of scattering is angularly unresolved to the gravitational lens but of similar scattering time to the gravitational delay.

\begin{figure}[!htb]
    \centering    
    \includegraphics[width=\columnwidth]{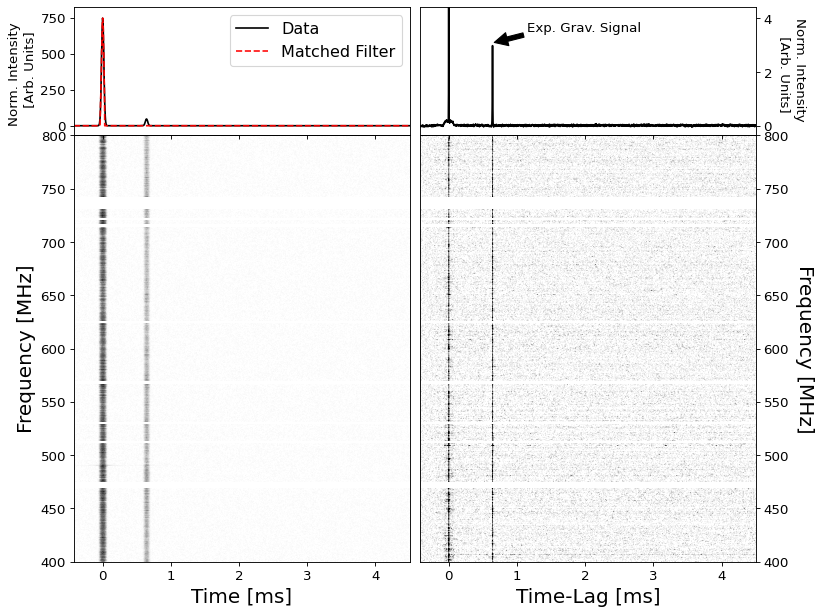}
    \caption{ Simulated waterfall (left) and time-lag correlation (right) of lensing from a scattering screen and a 10 solar mass gravitational lens. The scattering screen produces diffractive scintillation is manifested as an interference pattern in the spectra of the FRB along the frequency dimension shown in the left panel. The phase correlation search produces a correlation signal at the time delay expected for a gravitational lens. Frequencies with caustics are masked. } 
    \label{fig:grav_weak_gscatt_dual}
\end{figure}

The simulated baseband data for scenario 1 is shown in figure \ref{fig:grav_weak_gscatt_dual}. In the left panel, the small scattering time produces the diffractive scintillation seen in the spectra of the first image. In the right panel, the time-lag correlation shows the phase correlation is being preserved as the scattering is too small to alter the phase between the gravitationally lensed images. In this scattering limit, the system can be considered coherent and unresolved.

\begin{figure}[!htb]
    \centering    
    \includegraphics[width=\columnwidth]{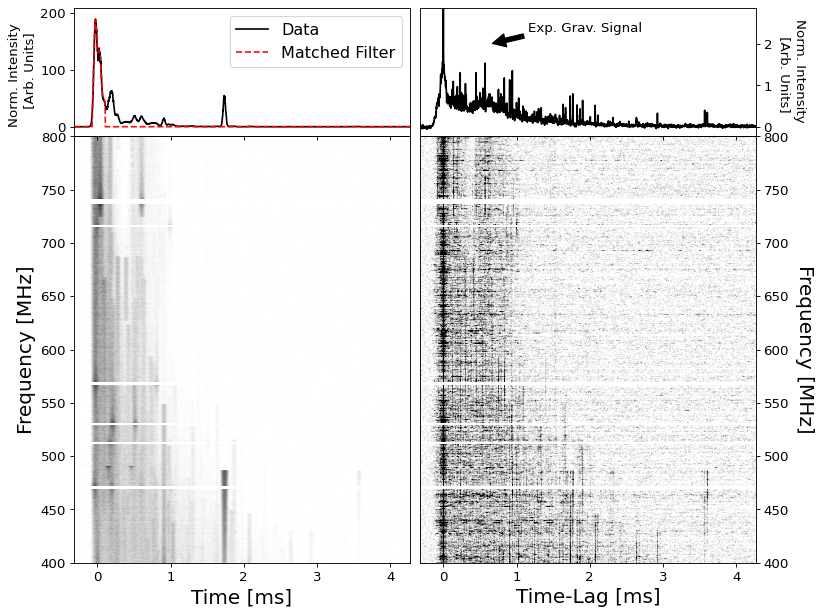}
    \caption{ Simulated waterfall (left) and time-lag correlation (right) of lensing from a strong scattering screen and a 10 solar mass gravitational lens. The scattering screen produces images over a timescale comparable to the gravitational time delay, shown in the left panel. The phase correlation search produces no detectable correlation signal at the time delay expected for a gravitational lens. In this limit, the scattering has removed the coherence between the gravitational images. Frequencies with caustics are masked.} 
    \label{fig:grav_strong_gscatt_dual}
\end{figure}

In scenario 2, we increased the strength of scattering by adjusting $\sigma_{\mathrm{DM}}$ through $\tau_{\mathrm{scatt}}$. The resulting baseband simulation is shown in figure \ref{fig:grav_strong_gscatt_dual}. This system has a gravitational time delay comparable to the scattering time. This produces a morphology, shown in the left panel, where the second image is only visibly distinguishable at higher frequencies. The time-lag correlation shows that, while the scattering screen creates images that collectively produce excess correlation power, the gravitational image is undetectable as an achromatic signal at one delay. The expected signal position is shown by the arrow in the top-right panel. In this limit, the system is considered incoherent and the gravitational lens is ``resolving" the screen.

\begin{figure}[!htb]
    \centering    
    \includegraphics[width=\columnwidth]{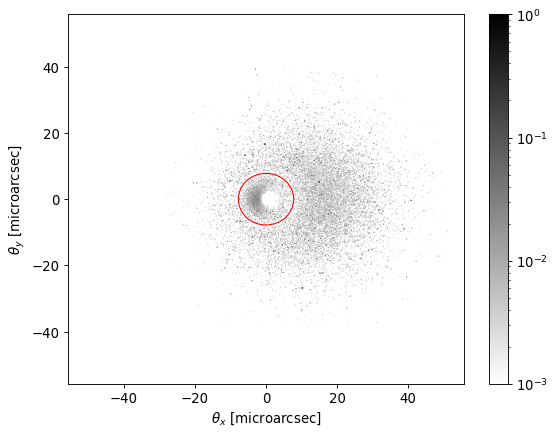}
    \caption{ Simulated angular position of images generated for a 400 MHz radio wave propagated through a scattering screen and then a 10 solar mass gravitational lens. The scattering screen is considered to be ``resolved'' by the gravitational lens as the angular broadening from scattering is comparable in scale to the angular extent of the gravitational lens. The color bar indicates the magnification of each image.} 
    \label{fig:grav_spatial_resolved}
\end{figure}

Figure \ref{fig:grav_spatial_resolved} shows the spatial distribution of images generated from this scenario. In this case, the scattering time is comparable to the gravitational delay. The angular extent is of a similar scale to the gravitational lens so we consider the scattering screen to be ``resolved" by the gravitational lens. The scattering screen is producing the speckle distribution of images while the gravitational lens is causing the secondary warped image to appear within the Einstein radius. In other words, while one could see the gravitational lens in spatial coordinates a time-lag search shows the system has lost its phase coherence. In this instance, without resolving the spatial distribution of images, an observer cannot assert whether the second intensity peak is a lensed copy of the first peak or merely a second burst from a repeating FRB with similar intensity.

\begin{figure}[!htb]
    \centering    
    \includegraphics[width=\columnwidth]{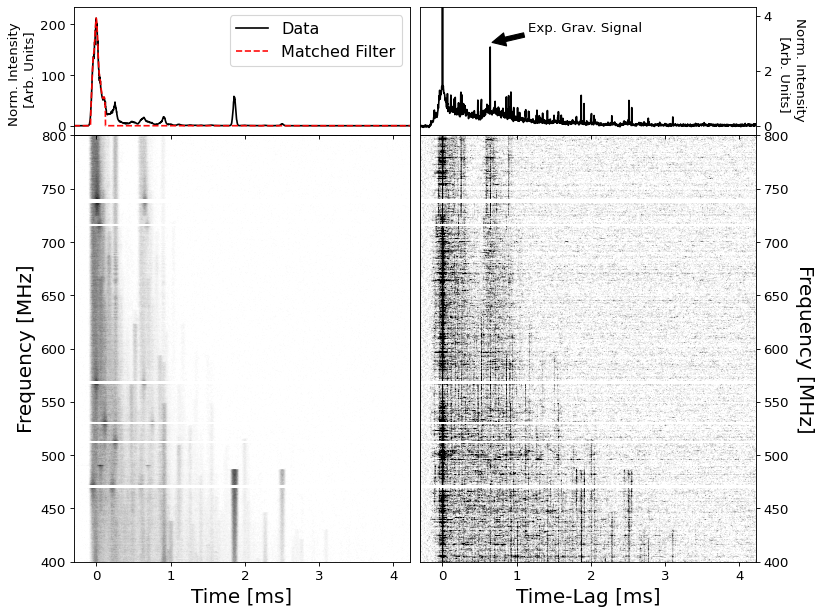}
    \caption{ Simulated waterfall (left) and time-lag correlation (right) of lensing from a strong scattering screen and a 10 solar mass gravitational lens. The scattering screen produces images over a timescale comparable to the gravitational time delay, shown in the left panel. The phase correlation search finds a sharp correlation signal at the time delay expected for the gravitational lens. In this limit, even though the timescales are similar, scattering has not removed the coherence between the gravitational images. Frequencies with caustics are masked.} 
    \label{fig:grav_strong_unres_gscatt_dual}
\end{figure}

In scenario 3, we preserve the timescale of scattering but place the scattering screen to be 1 kpc from the FRB. The total extent of scattering is in the regime that it remains a point-like source to the gravitational lens. The simulated baseband for this scenario is shown in figure \ref{fig:grav_strong_unres_gscatt_dual}. Similar to figure \ref{fig:grav_strong_gscatt_dual} the two distinct images are seen at higher frequencies in the left panel. Unlike the right panel of figure \ref{fig:grav_strong_gscatt_dual}, the right panel of figure \ref{fig:grav_strong_unres_gscatt_dual} has a distinct phase coherent signal present at the expected gravitational delay. The phase is shown to be preserved as the angular extent of the scattering screen is still in the regime where it is considered a point-like source to the gravitational lens. However, note the coherent signal present here is only detectable due to a prior expectation. For a real event, if the magnification of the gravitational lens was not large enough to be significantly above the correlation power of the scatter, it would be undetectable. Nonetheless, coherence is preserved in this scenario.

From these different examples, the simulation produces two implications for the observation of phase coherent gravitational lensing, the first implication is that, without using any spatial information, a time-domain search allows for the detection of a gravitational lens if the angular extent of scattering is point-like to the gravitational lens. This follows the expectation present in section \ref{sec:coherence}. For coherence to be held propagating through a single screen, the angular extent of the previous screen must be point-like. For gravitational lensing timescales smaller than the burst timescales, one expects a null response if the scintillation washes out the delay of the lens \cite{Katz2020}. This statement relies on two physical assumptions being satisfied, (a) the FRB is a point source to the scattering screen and (b) the lensing system does not change over the delay time between the gravitational images. The second implication is that when considering the effects of scattering to search for a gravitational lens, searching with smaller frequency channel resolutions is preferable. The average phase change over the channel needs to be smaller than the phase imprinted by a lens for a coherent phase correlation to be preserved.

In \textcite{Katz2020}, the authors considered the effects of scintillation in the limit where the gravitational time delay is smaller than the width of the FRB burst. They found scattering screens should decohere the interference pattern of a lensing signal in the FRB spectrum for masses above $0.1 M_{\odot}$. In this example, we have shown that for masses above $0.1 M_{\odot}$ coherence can still be maintained even if the scattering time is comparable to the gravitational time delay through scenario 3. Our simulation also shows the main concern for the coherence of a gravitational lens is whether the angular extent of scattering is comparable to the coherence scale of the gravitational lens. The worst-case scenario would be searching for phase coherent microlensing in a galaxy. Two screens close to each other would increase the chances that the angular broadening from scattering be comparable to or larger than the scale of any gravitational lens. In \textcite{Kumar2023}, the authors presented the case where scattering and gravitational lensing occur on or near the same plane, physically motivated by microlensing of an FRB due to objects embedded in an intervening galaxy. The authors find that, below masses of $<100 M_{\odot}$, the signal should be suppressed. In this work, we consider the phase coherence of the images themselves and show similarly that if the scattering extent and gravitational lens are of similar scale lensing can be suppressed. However, it should be noted that a phase coherent search of voltage data is more sensitive than intensity alone \cite{Kader2022}, meaning one can put limits on how point-like the screens might be, i.e. is a system represented by scenario 2 or scenario 3. The only problem, in this case, occurs when the achromatic delay response is decohered such that the gravitational lens would be degenerate with a plasma lens.

For the use of FRBs to probe the universe for compact dark matter, especially between the $1-10 M_{\odot}$ region of interest for primordial black holes (PBHs) \cite{Munoz2016}, this suggests that FRBs are more suitable to probe for objects largely disconnected from any intervening galaxy since PBHs in galaxies suffer from decoherence. As a final remark, it should be noted that, even if phase coherence is lost for larger masses, the gravitational lensed FRB can still be identified from the spatial distribution of images, e.g. figure \ref{fig:grav_spatial_resolved}, if one could spatially resolve these images better than a microarcsecond resolution.

\section{Conclusion}

In this paper, we have provided an overview of lensing formalism for radio wave propagation relevant to compact emission sources such as FRBs and pulsars and created a simulation tool to generate the phase preserving morphologies of FRBs from these propagation effects. The tool works assuming a semi-classical approximation for the propagation and that the lensing is represented by a thin screen. In this paper, we have used this tool to generate possible morphologies for gravitational lensing, refractive plasma lensing, and interstellar plasma scattering of radio waves. We created multi-plane morphologies by showing the combined propagation for some example configurations. For these cases, we presented the morphologies and time-lag correlation signatures to inform observational search algorithms and non-detection probability calculations.

We compared the simulation code results with an analytically calculable case consisting of a point mass gravitational lens and a rational plasma lens. We found the simulation has at worst a fractional error of $10^{-2}$ for grids of order $1000 \times 1000$ points, the typical grid size with reasonable ($<$ day) compute times. The scaling of fractional error with grid size is shown in figure \ref{fig:sim_error}. Figure \ref{fig:sim_error_freq} shows the simulation has a fractional error that can be considered independent of frequency.

One remaining area of potential improvement for the simulation is how caustics are dealt with. As the simulation assumes the geometric optics regime, images near or at caustics will have asymptotic magnification that tends to infinity. This is easily identifiable and can be flagged, which was done in this paper. We leave for future work improvements or changes to the simulation algorithm that might account for this. The simulation can model morphologies relatively quickly ($\sim$minutes) such that we foresee the extension of this simulation to forward model observed FRB morphologies to fit the parameters associated with lensing systems. Another improvement to this simulation that is left for future work is the inclusion of polarization and magnetic fields to simulate birefringent lensing.

The work presented in this paper seeks to present a method to aid in understanding the changes to the morphology of FRBs from different propagation effects. Forward modeling the propagation effects present in measured signatures, such that they can be deconvolved from the observations, could remove the propagation effects, revealing the underlying physics of the FRB emission mechanism. The simulation could also be used to test different gravitational lensing models. In this paper, we presented one example case of a gravitational lens and a scattering screen and demonstrated how, depending on the timescale of scattering, one can decohere the phase response between two gravitationally lensed FRB images. One extension to this study could be to consider the effects of lensing by a galaxy which would include both plasma and gravitational contributions.

Another example case presented was a dual scattering screen system, where the two screens produced separate morphological changes to the FRB. In this example, the Milky Way screen created diffractive scintillation while the FRB host screen created a scattering tail. In this paper, we restricted ourselves to a Gaussian process to generate the scattering but we note that the simulation could be used to test other models to generate the scattering morphology. It allows for an exploration into the lensing parameter space that could produce a scattering phenomenon.

In summary, this simulation toolset provides an opportunity to test whether the observed FRB morphology is intrinsic or a propagation effect. This toolset also allows one to investigate which lensing systems can create coherent lensing and what their time-lag phase response would be. With an increasing number of FRBs being observed, the chances of observing a lensing event, whether from plasma or a gravitational lens, also increase. This toolset can allow us to be prepared for this eventuality and optimize our search methods to detect and distinguish these lensing events. 

\begin{acknowledgments}
We thank Robert Main for their useful comments, suggestions, and references on scintillation.
\allacks
\end{acknowledgments}

\bibliographystyle{apsrev4-2}
\bibliography{main.bib}

\end{document}